%% file: main.tex
\documentclass[a4paper]{article}
\usepackage[english]{babel}
\usepackage{longtable}
\usepackage[utf8]{inputenc}
\usepackage[aboveskip=1pt,labelfont=bf,labelsep=period,justification=raggedright,singlelinecheck=off]{caption}

\usepackage{float}

\usepackage{color,soul}

\usepackage{url}
\usepackage{hyperref}
\usepackage[style=numeric,backend=biber,sorting=none]{biblatex} 
\input{preamble}

\title{Projecting Multimorbidity and Mortality under Demographic Change and Preventive Interventions}
 
\author{Katharina Ledebur$^{1,3,4}$ \and Alexandra Kautzky-Willer$^{2}$ \and Stefan Thurner$^{1,3,5}$ \and Peter Klimek$^{1,3,4*}$}

\date{
	$^1$Institute of the Science of Complex Systems, CeDAS, Medical University of Vienna, Spitalgasse 23, A-1090 Vienna, Austria\\%
	$^2$Gender Medicine Unit, Division of Endocrinology and Metabolism, Department of Internal Medicine III, Medical University of Vienna, Spitalgasse 23, A-1090 Vienna, Austria\\%
    $^3$Complexity Science Hub Vienna, Metternichgasse 8, A-1030 Vienna, Austria\\
    $^4$Supply Chain Intelligence Institute Austria, Metternichgasse 8, A-1030 Vienna, Austria\\
    $^5$Santa Fe Institute, 1399 Hyde Park Road, Santa Fe, NM 87501, USA\\
    $*$\texttt{peter.klimek@meduniwien.ac.at}\\
}

\begin{document}
\maketitle

\bigskip

\begin{abstract} 

As populations age, the rise of multimorbidity poses a significant healthcare challenge. However, our ability to quantitatively forecast the progression of multimorbidity remains limited. Leveraging a nationwide dataset comprising approximately 45 million hospital stays spanning 17 years in Austria, we develop a new compartmental model for chronic disease trajectories across 132 distinct multimorbidity patterns (compartments). Each compartment represents a distinct constellation of co-occurring chronic conditions, with transitions modeled as age- and sex-dependent probabilities. We use the compartmental disease trajectory model (CDTM) to simulate disease trajectories to 2030, estimating the frequency of all empirically observed co-occurrence patterns among more than 100 diagnosis groups. We demonstrate the model's utility in identifying high-impact prevention targets. A 5\% reduction in new cases of hypertensive disease (I10--I15) leads to a 0.57 (SD 0.06)\% reduction in all-cause mortality over a 15-year period, and a 0.57 (SD 0.07)\% reduction in mortality for malignant neoplasms (C00--C97). We also evaluate long-term impacts of SARS-CoV-2 sequelae, projecting earlier and more frequent hospitalizations across a range of diagnoses. Our fully data-driven modelling approach identifies leverage points for proactive preparation by physicians and policymakers to reduce the overall disease burden in the population, emphasizing patient-centered healthcare planning in aging societies. \\           

\end{abstract}

\maketitle

\section{Introduction} \label{sec:intro}

In many Western countries, life expectancy has steadily increased over the past two decades~\cite{world_health_organization_global_nodate}. Across Europe, it rose by 5.8 years between 1999 and 2019, while in Austria it increased by 3.1 years for men and 4.5 years for women~\cite{our_world_in_data_life_nodate, statistics_austria_healthy_2024}. However, this progress masks an emerging paradox. In Austria, the average number of years lived in (very) good health has declined by 2.8 years for men and 1.9 years for women between 2014 and 2019~\cite{statistics_austria_healthy_2022}. As populations age, the prevalence of chronic diseases increases, leading to longer periods lived with illness~\cite{soh_morbidity_2020, raleigh_trends_2019}. These shifts reflect medical advances that prolong life, but also expose a growing gap in the prevention and management of noncommunicable diseases (NCDs)~\cite{bachner_health_2018}. 

As chronic diseases often share risk factors, this increased prevalence of chronic diseases is expected to increase the number of people with multiple health conditions \cite{whitty_rising_2020}. This phenomenon, referred to as "multimorbidity", is one of the major challenges facing countries with aging populations \cite{pearson-stuttard_multimorbiditydefining_2019}.  

Most existing research on multimorbidity uses highly aggregated comorbidity indices such as Charlson or Elixhauser~\cite{johnston2019defining,charlson1987method, elixhauser1998comorbidity} or focuses on specific conditions or populations \cite{marengoni_prevalence_2008, zhang_prevalence_2019, van_den_bussche_which_2011, kingston_projections_2018, Althoff2024, fortin_rperseearvchaalrteicnle_2010, pefoyo_increasing_2015, divo_ageing_2014}. 

One reason for the limited understanding of how multimorbidity develops and progresses over time is the lack of a consistent and dynamic definition of multimorbidity \cite{nicholson_measurement_2019}. Definitions of multimorbidity often rely on diagnosis counts, despite evidence that health outcomes depend strongly on the specific combination and order of conditions~\cite{chmiel_spreading_2014, pearson-stuttard_multimorbiditydefining_2019}. An alternative to defining multimorbidity by the count of diagnoses is to define it as sequences of diseases over the life course, known as disease trajectories~\cite{haug_high-risk_2020}. 

Numerous studies of phenotypic comorbidity networks have advanced the field of network medicine and our understanding of disease trajectories. These networks show diseases (nodes) and their associations (links) with factors like genetics, lifestyle, or environmental exposures. Population-wide electronic health records (EHRs) and medical claims data have also allowed us to identify these trajectories on a large scale \cite{chmiel_spreading_2014, hidalgo_dynamic_2009,  jensen_temporal_2014, jeong_network-based_2017, fotouhi_statistical_2018, roque_using_2011, hu_large-cohort_2019, haue_temporal_2022, beck_diagnosis_2016, giannoula_identifying_2018, giannoula_system-level_2020, giannoula_identifying_2023}. Patients tend to develop diseases that are close in network space to their existing conditions. This facilitates predicting future diseases based on medical history \cite{hidalgo_dynamic_2009, chmiel_spreading_2014}.

A major challenge in studying long-term disease trajectories is the large number of possible combinations of diseases and their temporal permutations. Clustering frequently co-occurring diseases and quantifying how they influence each other can reliably uncover long-term disease trajectories as a series of disease combinations (health states) for populations within countries \cite{haug_high-risk_2020, savcisens_using_2024}. However, studies on the temporal evolution of multimorbidity, especially in the context of an aging population, remain scarce, both in terms of retrospective data analyses \cite{van_oostrom_multimorbidity_2012, uijen_multimorbidity_2008, harrison_prevalence_2016}, and for predictive modelling. The bottlenecks are not only data requirements but also methodological challenges \cite{kingston_forecasting_2018, kingston_projections_2018}.

More specifically, there is a long-standing tradition of using mathematical models to predict the spread of infectious diseases. They played a pivotal role in decision-making during the Covid-19 pandemic by modelling the potential impact of public health interventions. Mathematically, the future prevalence of infectious diseases can be modelled using contagion processes on networks. Chronic non-communicable diseases, on the other hand, emerge from the interaction of numerous disease-causing mechanisms ranging from genetic predisposition to lifestyle factors. Therefore, it is unclear whether accumulation of multiple chronic conditions in aging individuals as a can be understood as a network spreading process.

We address this gap by developing a compartmental model, based on a multimorbidity network to investigate multimorbidity based on chronic disease trajectory dynamics within the Austrian population. Our starting point is the observation that many chronic diseases tend to co-occur due to a shared etiology. 
We identify groups or "multimorbidity clusters" of frequently co-occurring diagnoses and quantify the likelihood that patients will progress from one cluster to another over time. 

\begin{figure*}  
    \centering

    \captionsetup{width=.9\textwidth}
    \includegraphics[width=.9\textwidth]{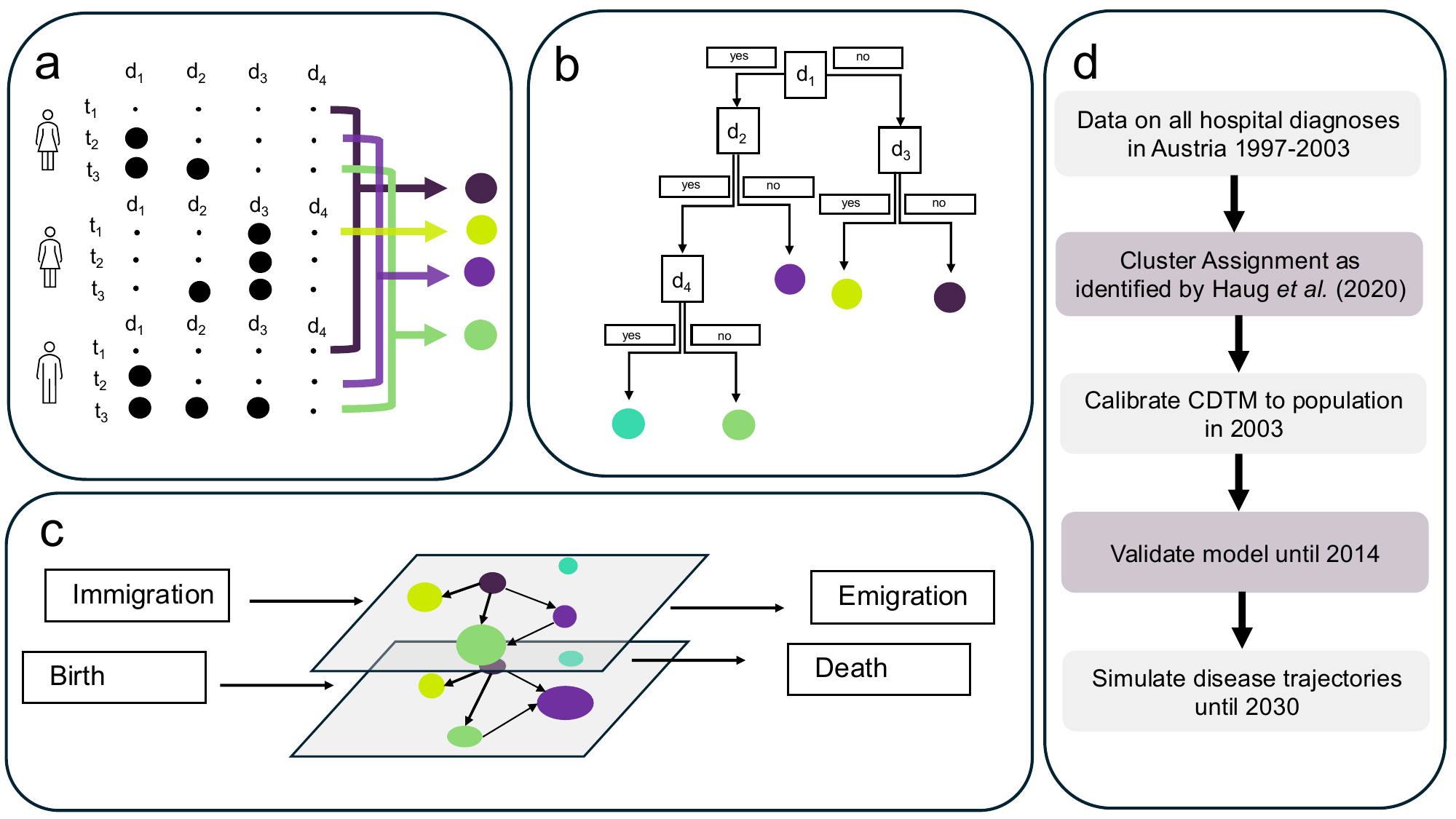}
    \caption{A data-driven compartmental model of chronic disease trajectories.
    (a) Each row represents the health state of a patient at a specific year $t_j$, encoded as a binary vector across diagnosis dimensions $d_i$. A large dot indicates presence of a diagnosis in that year. These vectors are clustered using DIVCLUS-T, a hierarchical divisive algorithm, assigning each patient-year to a disease cluster (colored arrows), which defines compartments in the CDTM. (b) DIVCLUS-T builds a binary decision tree based on diagnosis presence/absence, minimizing intra-cluster variance. Each internal node checks for a diagnosis (e.g., “has $d_1$?”), branching accordingly. Leaf nodes represent clusters; each one contains a set of patients with similar diagnostic patterns.
This tree allows fast and interpretable assignment of new observations to clusters, as a sequence of binary questions. (c) The CDTM simulates population health dynamics using annual processes of birth, death, immigration, and emigration. Individuals move between 132 disease clusters based on age ($a$)- and sex ($g$)-stratified  transition probabilities ($q'_{g, a, k, j}$) , calibrated on past transitions. (d) We initialize the model using the diagnostic histories of all individuals who were alive between 1997 and 2003. Patients are then assigned to multimorbidity clusters based on their diagnoses. We calibrated the model to Austria’s 2003 population, validated it until 2014, and used it to simulate disease trajectories until 2030.
    }
    \label{fig:vis_abstract}
\end{figure*}

While for infectious disease dynamics compartments typically correspond to e.g. susceptible, infected, or recovered individuals, for chronic diseases these compartments correspond to different multimorbidity clusters (Figure\ref{fig:vis_abstract}a) and every patient is uniquely assigned to one of these clusters, or "health states" (Figure\ref{fig:vis_abstract}b).

These clusters are constructed using a hierarchical clustering algorithm applied to data from approximately 45,000,000 hospital stays (covering about 9,000,000 individuals in Austria between 1997 and 2014), as previously proposed \cite{haug_high-risk_2020}. 
The method identified 132 distinct disease clusters that serve here as the basis for a new population-scale compartmental model. The CDTM framework is outlined in Figure \ref{fig:vis_abstract}c. Each compartment corresponds to a multimorbidity cluster, and transitions are modeled as discrete-time probabilities based on age and sex. The model simulates the health trajectories of the entire Austrian population from 2003 to 2030. It incorporates demographic dynamics, such as births, deaths, and migration, based on national statistics. The model also allows individuals to acquire new diagnoses based on their current health state. Capturing the interacting dynamics of disease accumulation is essential for projecting future health outcomes. This framework supports in-sample reconstructions and out-of-sample projections of the development of diseases and their combinations across the entire diagnostic spectrum at the population level. We demonstrate its usefulness through three types of analyses.
 
We apply the model to (1) project the annual incidence trends of all ICD-10 diagnoses (A00–N99) and their combinations under a baseline scenario that assumes constant transition dynamics and a changing population structure, (2) examine the system-level impact of hypothetical prevention scenarios involving reduced probabilities of specific diagnoses on population-wide health outcomes, and (3) evaluate the long-term impact of post-acute SARS-CoV-2 sequelae on hospitalization trajectories across diagnoses.

\section{Methods} \label{sec:methods}
\subsection{Data Characteristics and Compartmental Disease Trajectory Model (CDTM)}
The CDTM is based on a comprehensive dataset provided by the Austrian Ministry of Health covering all hospital records in Austria from 1997 to 2014. It includes approximately 45 million hospital stays for over 9 million individuals. Each record contains patient sex, 5-year age group, length of stay, discharge reason (e.g., discharge, transfer, or death), and primary and secondary diagnoses coded at the 3-digit ICD-10 level.

To identify the multimorbidity clusters, we use a previously proposed approach using the DIVCLUS-T hierarchical clustering algorithm applied to patient diagnosis trajectories~\cite{haug_high-risk_2020}.  Clustering is performed on individuals with no recorded hospital diagnoses from 1997–2002 who received at least one diagnosis after 2003. Each cluster represents a distinct pattern of co-occurring conditions, defined by inclusion and exclusion criteria. These clusters serve as compartments in the CDTM (Figure \ref{fig:vis_abstract}b). For a detailed description of the clusters see section~\ref{sec:cluster_criteria} in the supplementary information.

For each age-sex stratum, we compute a transition matrix $q_{g,a,k,j}$ as the probability of individuals moving from cluster $k$ to cluster $j$ within a single time step based on their age $a$ and sex $g$. These probabilities are derived from observed transitions in the dataset by calculating the frequency with which individuals in cluster $k$ at time $t$ move to cluster $j$ at time $t+1$. 

We initialize the model by assigning individuals with hospital stays between 1997 and 2003 to a cluster based on their diagnosis history. To match Austria’s official population distribution at the start of 2003, we add the difference from census data into a "healthy" (diagnosis-free) compartment (cluster 0), representing individuals without hospital diagnoses (Figure \ref{fig:vis_abstract}a) \cite{statistics_austria_population_2024}.

Each patient's health status at the end of each year is assigned to a cluster based on their recorded diagnoses.

Each cluster $c$ is characterized by a set of 131 probabilities $p_{c,d}$, which describe the probability of receiving a diagnosis in group $d$ when transitioning to cluster $c$, where inclusion diagnoses satisfy $p_{c,d} = 1$, exclusion diagnoses satisfy $p_{c,d} = 0$, and all other diagnoses have probabilities in the range $0 \leq p_{c,d} \leq 1$. These probabilities are derived from the distribution of diagnoses within clusters of the modeled Austrian population of 2003.

From 2003 onward, we model individuals as probabilistically transitioning between disease clusters in a network, where transition probabilities depend on age and sex. For an exemplary network for females from 50 to 59 years old see Figure~\ref{fig:mmnetwork} in the supplementary information. In each simulation year, an individual of sex ($g$) and age ($a$) transitions from cluster $k$ to cluster $j$ with probability $q_{g,a,k,j}$. Upon entering a new cluster, diagnoses are probabilistically assigned based on $p_{j,d}$. This structure enables yearly projections of diagnosis incidence and multimorbidity burden across the entire population.

\subsection{Calibration and Simulation}
The model accounts for demographic changes. Each year individuals are born, migrate, and pass away with rates taken from historical data and a baseline scenario of future population movements provided by Statistics Austria \cite{statistics_austria_population_2024}. 
Mortality rates by cause are available until 2019; from that point onward, we assume that the distribution of deaths by cause, sex, and age remains constant at 2019 levels.

We calibrated the transition probabilities between multimorbidity clusters over the period 2003–2014 in a two-step process. First, we make a naive estimate of $q_{g,a,k,j}$ by measuring the frequency of transitions between clusters $k$ and $j$ in age group $a$ and sex $g$ directly in the data. We then rescale these values to ensure that the rate at which diagnoses are acquired in the simulation aligns with empirical observations. We optimized via grid search to match the cumulative diagnoses per individual in the model to observed values. This led to an effective reduction of the naive first order estimates for $q$ by 65\% for individuals aged 0-29, 60\% for 30-39, 40\% for 40-49, 20\% for 50-59, and 10\% for 60-69. There is no reduction applied for people 70 and older. Figure~\ref{fig:mminc}a-d shows the number of diagnoses per individual across 10-year age groups for both the dataset and the model for the periods 2004-2006 and 2012-2014.

Second, we compute the discrepancy $\delta d(a)$ between observed and simulated diagnosis counts by diagnosis, age, and sex group. To reduce noise in age groups with low discrepancies, we merge the first three ten-year age groups into a single category (0-29 years) and combine the 30-39 and 40-49 age groups. The remaining age groups are kept at 10-year intervals. We distribute $\delta d(a)$ annually across age–sex groups to refine disease dynamics. Figure~\ref{fig:cal} shows the comparison of incidence rates for 2014 from the dataset and the model.
 
For the projections beyond 2015, we use a simplified version for the second step. We calculate a fixed rate adjustment for each diagnosis, age, and sex group, using the average observed rates from 2012 to 2014 as a basis. These adjustments remain constant throughout the projection period, ensuring the model remains consistent with recent trends while allowing for changes that reflect demographic shifts and prevention scenarios.
\subsection{Observables}
We define several outcome measures to characterize the simulated dynamics of chronic disease and multimorbidity.\\
\textbf{Relative incidence change.}\\
To measure how the incidence of a diagnosis group changes over time, we compute the relative incidence change: 
\begin{equation}
    \Delta I_{rel}  = \frac{I_{y+\Delta t} - I_{y}}{I_{y}} 
\end{equation}
which quantifies by how much the incidence of a particular diagnosis has changed in year $y + \Delta t$, relative to the initial incidence in year $y$.\\
\textbf{Multimorbidity burden.}\\
We assess multimorbidity at two levels: (i) as \textit{multimorbidity incidence} which refers to the frequency of individuals receiving $n$ diagnoses ($n$ = 1-3, 4-6, 7-9, >10) and (ii) as \textit{relative occurrence} of a multimorbidity cluster, $c$, per year $y$, defined as the ratio of transitions to cluster $c$, to the total number of transitions in that year.\\
\textbf{Mortality.}\\
The decrease in all-cause mortality (total number of deaths) due to reduction of new-onset diagnoses is measured in \textit{reduction in death numbers} and \textit{prevented deaths per prevented diagnosis}. We measure the \textit{reduction in death numbers} by comparing the total number of deaths in a simulation run where the new-onset diagnoses of a particular diagnosis group is reduced by a factor, $\gamma$, to deaths due to all death causes in a simulation with no reduction. \textit{Prevented deaths per prevented diagnoses} are computed by dividing the total number of prevented deaths by the total number of reduced diagnoses for the simulation where a certain diagnosis group is reduced by a factor $\gamma$.\\
\textbf{Covid-19 Shock.}\\
The impact of the Covid-19 Shock on the overall disease burden is estimated by computing the difference in number of diagnoses from 2022 to 2030 between the main and the respective Covid-19 Shock scenario, compared to the main scenario (\textit{difference in incidence}). Additionally, we compute the \textit{average years until diagnosis} and compare the Covid-19 Shock to the main scenario. This quantifies how much earlier on average individuals are diagnosed with a diagnosis group as a result of the simulated Covid-19 Shock. 

\subsection{Scenarios}
In addition to the baseline scenario, we present scenarios that demonstrate the versatility of the developed CDTM. These scenarios explore how changes in diagnosis incidence - both reductions and increases - affect the health status of the population.

To assess the potential impact of interventions such as prevention efforts or the introduction of new medications, we modify the transition probabilities in the scenario simulations. Specifically, we reduce the incidence of a single diagnosis $d$ from the 131 diagnostic groups by adjusting the transition probability $q_{g,a,k,j}$ using a multiplicative factor $\gamma$ (where $\gamma = 0.95$): \begin{equation}\label{eq:reduction} \begin{split} q_{g,a,k,j}' = (1 - p_{j,d}) q_{g,a,k,j} + \ p_{j,d} p_{k,d} q_{g,a,k,j} + \ (1 - p_{k,d}) p_{j,d} q_{g,a,k,j} \gamma \quad. \end{split} \end{equation}

The first term in Eq.\ \eqref{eq:reduction} corresponds to transitions without the target diagnosis $d$, the second accounts for individuals already diagnosed with $d$, and the third reflects new diagnoses of $d$, adjusted by $\gamma$.

To estimate the potential impact of the Covid-19 pandemic on future disease incidence and mortality, we implement a "Covid-19 Shock" scenario. In this scenario, we assume that by 2022, every person in Austria has been infected with SARS-CoV-2 at least once. We then simulate the mortality associated with post-acute sequelae of SARS-CoV-2 infection based on empirical data~\cite{bmsgpk_covid-19_2024}.  
To estimate the increased likelihood of individuals being diagnosed with specific diseases following SARS-CoV-2 infection, we use the hazard ratios (HR) for specific sequelae reported in the retrospective cohort studies (see Table \ref{tab:cshr} for HRs and ICD-10 codes). We analyze the frequency of relevant ICD-10 codes codes (stratified by age and sex) within the 131 diagnostic groups in the original dataset. These frequencies are then scaled by the probability assigned to each diagnostic group. The resulting values are used to externally assign additional diagnoses within the population. Individuals are then reassigned to new multimorbidity clusters based on all acquired diagnostic groups. 

In both scenarios, decrease or increase in the annual death rates are resulting from the changes in annual diagnoses.

\section{Results} \label{sec:res}

\subsection{Impact of aging population on disease burden}
The out-of-sample simulation time span ranges from 2015 to 2030. During this period, we evaluate projected annual changes in the incidence of the 131 diagnostic groups, to explore how disease burden evolves in the population.

From 2015 to 2030, total incidence increased by 6.9\% (from 2.16 million to 2.31 million cases), while age-standardized incidence declined slightly, from 333,338 to 320,272 cases. This suggests that population aging is the primary driver of the absolute increase. The average relative error across diagnostic group estimates was 3.1\%.

\begin{figure*} 
    \centering
    \captionsetup{width=.95\textwidth}
    \includegraphics[width=.9\textwidth]{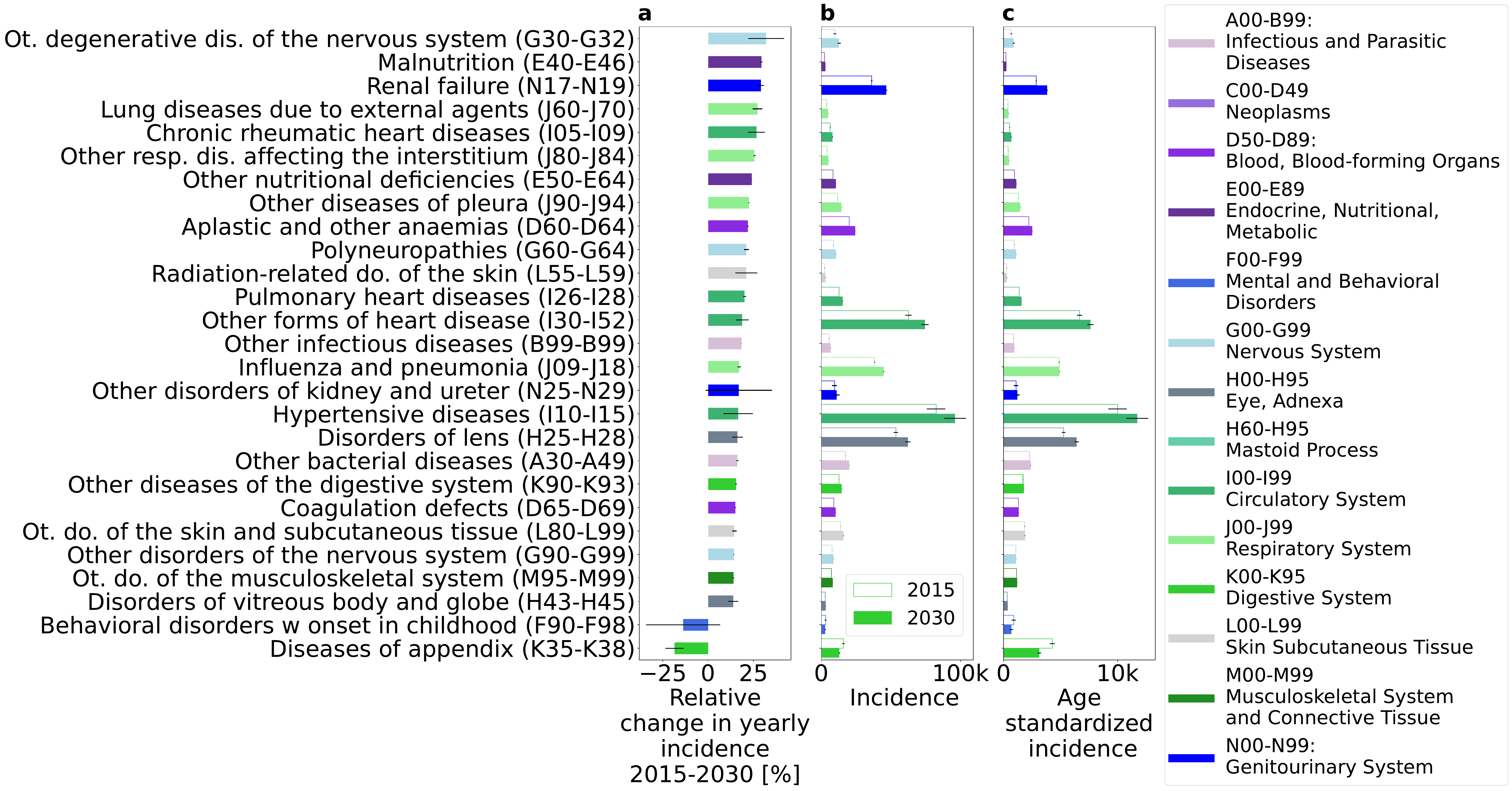}
    \caption{Projected changes in incidence of selected diagnosis groups. (a) Relative change in annual incidence of diagnosis groups with a relative absolute increase > 13 $\%$ 2030 compared to 2015 relative to 2015 and an incidence of more than 2000 in 2015. (b) Absolute values of annual incidence for 2030 and 2015. (c) Age standardized incidence in 2015 and 2030, assuming the 2015 population structure. Incidence increases are most pronounced for diseases of the nervous, genitourinary, and circulatory systems, as well as malnutrition. The comparison of crude and age-standardized values shows that for some diagnoses (e.g., renal failure, N17–N19), changes reflect not only demographic aging but also a worsening health profile. In all panels, error bars indicate the mean relative error from 2010 to 2014. Only diagnosis groups with stable error trends (i.e., constant error rates over time) are shown. We exclude groups where the absolute error in 2014 exceeds 5\% of the incidence or 500 cases, or where error fluctuates more than 1.3-fold between 2010 and 2014. }
    \label{fig:inc}
\end{figure*}

Figure \ref{fig:inc}a shows the relative change in incidence ($\Delta I_{rel}$) from 2015 to 2030 for selected diagnoses ($\Delta I_{rel}$ > 13\%). Figure \ref{fig:inc}b presents the corresponding absolute incidence values for 2015 (unfilled bars) and 2030 (filled bars). Error margins in parentheses indicate the mean relative error from 2010 to 2014. Diseases affecting the nervous system, the genitourinary system and the circulatory system show an increase in incidence. For other degenerative diseases of the nervous system (G30-G32) we observe $\Delta I_{rel}$ = 32.04\% (9.92), from 9,700 (960) diagnoses in 2015 to 13,000 (1,300) in 2030. For malnutrition (E40-E46) we find an increase of 24.49\% (9.92), from 2,200 (13) diagnoses in 2015 to 2,800 (17) in 2030. For renal failure (N17-N19), $\Delta I_{rel}$ = 29.21\% (1.63), increasing from 36,000 (590) diagnoses in 2015 to 46,800 (760) diagnoses in 2030. These diagnoses predominantly affect older adults.
Whereas relative incidence ($\Delta I_{\text{rel}}$) for diseases of the appendix decreases by 18.50\% (5.00), from 15,800 (800) diagnoses in 2015 to 12,800 (640) in 2030. These conditions primarily affect younger individuals (for age-stratified trends, see Figure\ref{fig:incage}). 

Age-standardized incidence values are computed by applying the 2015 age distribution to the 2030 incidence estimates (Figure \ref{fig:inc}c). An increase in age-standardized incidence, such as that seen with renal failure (N17–N19), indicates that the increase in incidence shown in Figure \ref{fig:inc}a and b is not solely driven by demographic aging but also reflects a deterioration in population health. For example, renal failure rises from 2,900 (50) diagnoses in 2015 to 3,250 (50) in 2030.

The CDTM further allows us to examine the development of the frequency of combinations of diagnoses from each multimorbidity cluster. From 2015 to 2030, we observe a 2.7\% increase in transitions from clusters with three inclusion diagnoses to those with four.

\begin{figure*} 
    \centering
    \captionsetup{width=.95\textwidth}
    \includegraphics[width=.95\textwidth]{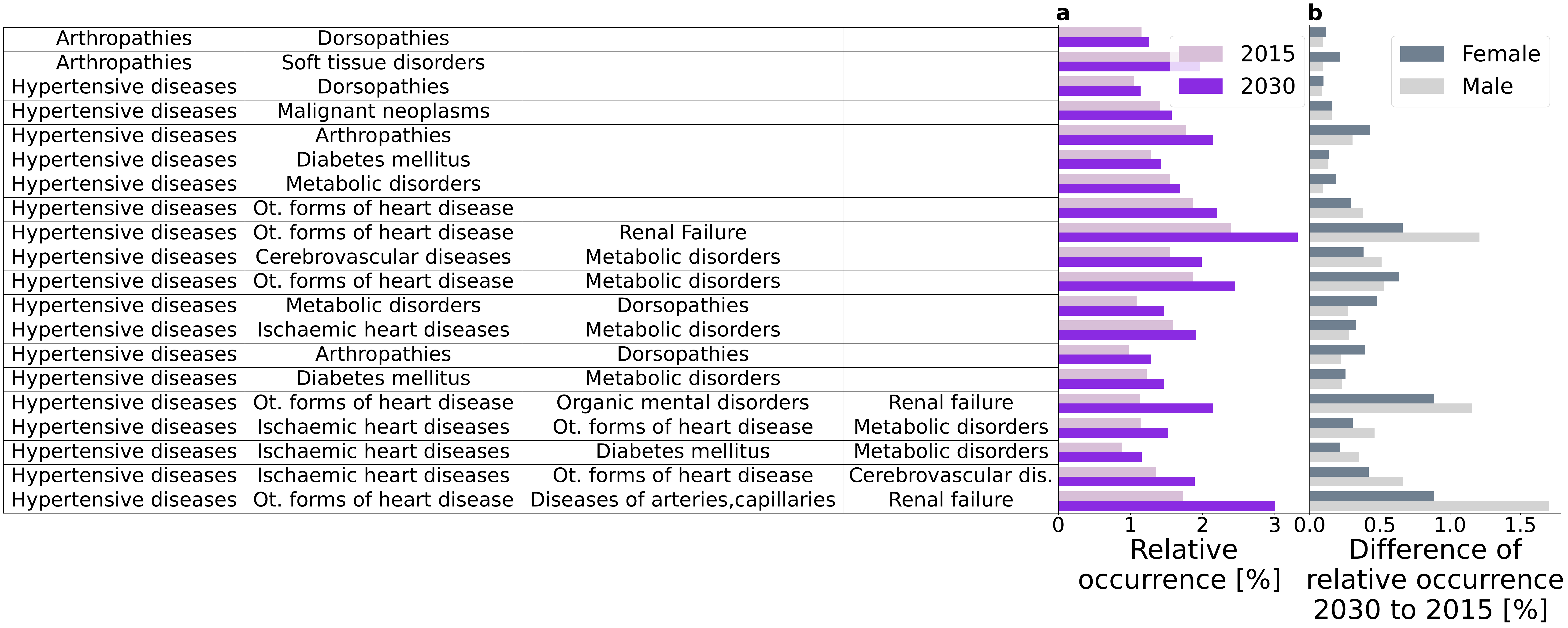}
    \caption{Relative occurrence of the 20 most common multimorbidity clusters in 2015. (a) Relative occurrence in the simulation year 2015 and 2030. We show the inclusion criteria of all 20 multimorbidity clusters. Especially clusters describing disease patterns containing hypertensive diseases (I10-I15) exhibit an increase between 2015 and 2030. (b) Difference in relative occurrence 2030 compared to 2015 for females and males, respectively. Differences in occurrence between males and females is only visible for multimorbidity cluster including diseases of the genitourinary system.}
   \label{fig:mminc_table}
\end{figure*}

Figure \ref{fig:mminc_table}a displays the 20 most common multimorbidity clusters in 2015, along with their relative frequencies in both 2015 and 2030. The most prevalent clusters include hypertensive diseases, a pattern that persists throughout the forecast period. For the disease pattern describing the co-occurrence of hypertensive diseases (I10--I15), other forms of heart disease (I30--I52), diseases of arteries, arterioles and capillaries (I70--I79) and renal failure (N17--N19) we find an increase in relative occurrence from 1.73\% to 3.0\% from 2015 to 2030. The relative occurrence of the pattern including hypertension, other forms of heart disease, organic, including symptomatic, mental disorders (F00--F09) and renal failure, increases from 1.13\% to 2.14\%. The third highest increase in relative occurrence from 2.40\% to 3.32\% is observed for hypertensive diseases, other forms of heart disease and renal failure. Figure \ref{fig:mminc_table}b contrasts the difference in this relative occurrence between 2030 and 2015, stratified by gender. The difference varies by gender mainly in disease patterns including renal failure. For instance, males show an increase of 1.70\% and females an increase of 0.88\% between 2015 and 2030 in the disease pattern encompassing hypertensive diseases, other forms of heart disease, diseases of arteries, arterioles, and capillaries, and renal failure. 

\subsection{Leverage points for optimal prevention schemes}

\begin{figure*} 
    \centering
    \captionsetup{width=.95\textwidth}
    \includegraphics[width=.8\textwidth]{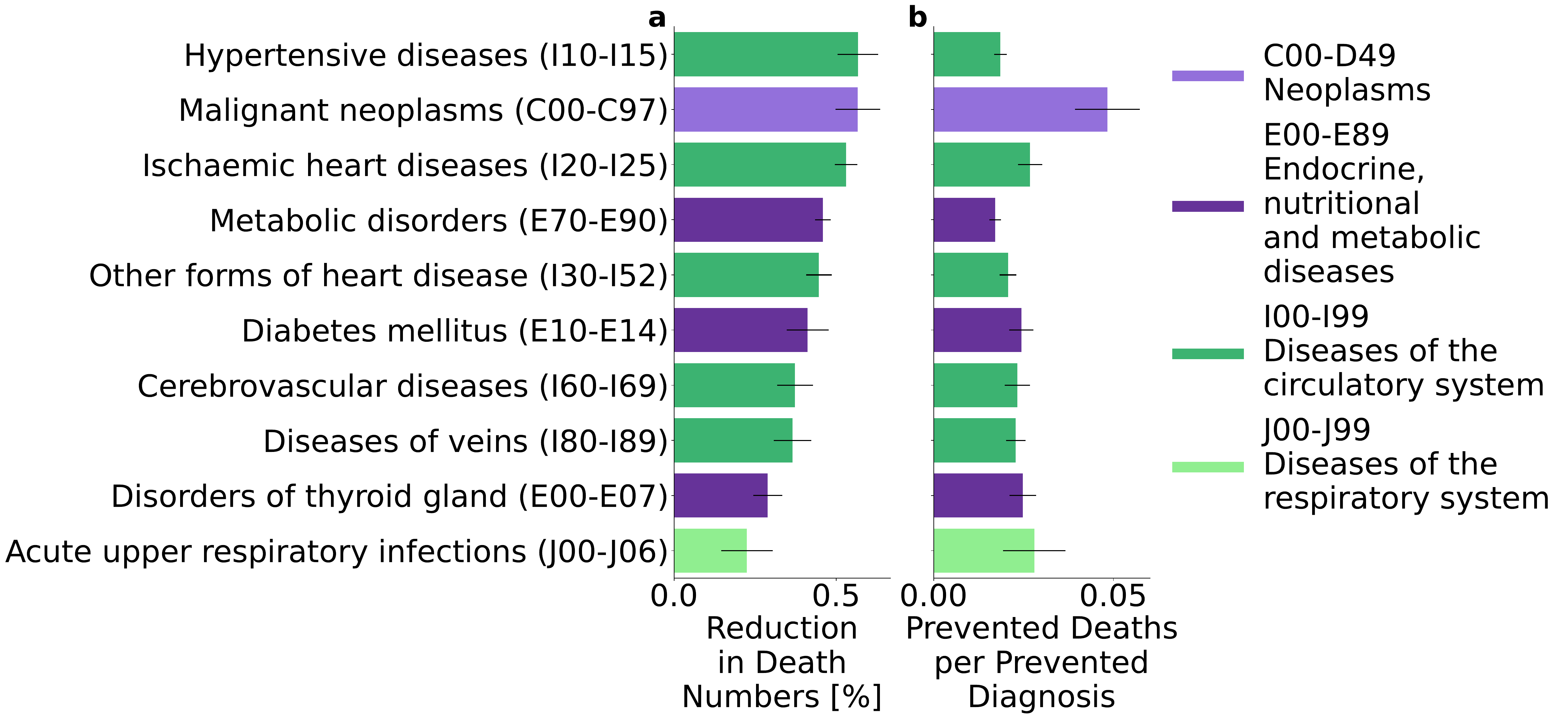}
    \caption{Impact of reducing disease incidence on overall mortality.(a) Reduction in all-cause deaths (2022–2030) following a 5\% annual decrease in transition probabilities of 10 diagnosis groups with largest effect on all-cause mortality. Per diagnosis group, $d$, on the y-axis we simulate a scenario where we annually reduce the transition probabilities by 5\% to clusters where the probability of being diagnosed with $d$ in the following timestep increases. (b) Ratio of number of prevented deaths per prevented diagnoses. To account for differences in the incidence of both of the reduced diagnosis and potential sequelae, we show the number of reduced deaths in relation to the number of reduced diagnoses (all diagnosis groups) from 2022 to 2030. Values represent the average of 10 simulations; error bars show the standard deviation.
    }
    \label{fig:red_deaths}
\end{figure*}

Targeted interventions can reduce disease incidence and downstream mortality. 
In the following, for each diagnosis, we evaluate how much a relative reduction of 5\% in the transition probabilities leading to new onsets of the diagnosis impacts all-cause mortality, see Figure \ref{fig:red_deaths}. We highlight the 10 diagnoses with the largest effect on all-cause mortality, focusing on potentially preventable conditions (e.g., those addressable through medication, prevention efforts, or vaccination; Figure \ref{fig:red_deaths}a). These groups include C00-D49, E00-E99, F00-F99, I00-I99, J00-J99 and N00-N99. Deaths prevented per diagnosis are highest for diseases affecting the circulatory system, neoplasms, endocrine, nutritional and metabolic diseases, and diseases of the respiratory system. Each bar shows the average 15-year reduction in deaths across 10 simulations in response to a 5\% decrease in the transition probabilities for individuals at risk of the corresponding diagnosis on the y-axis.

A 5\% reduction in new onsets of hypertensive diseases (I10–I15) is associated with a 0.57\% (SD 0.06) decrease in all-cause mortality. For malignant neoplasms, the reduction is 0.57\% (SD 0.07), and for ischaemic heart diseases (I20–I25), 0.53\% (SD 0.04). These conditions represent the leading causes of death in Austria \cite{statistics_austria_causes_2024}. Less common causes of deaths are diseases of veins (I80-I89) (0.36 (SD 0.06)\%) and disorders of the thyroid gland (E00-E07) (0.29 (SD 0.05)\%), suggesting that a reduction of new onsets of diseases in these diagnosis groups leads to a reduction in potentially fatal sequelae. 

To correct for potential biases related to differences in disease incidence and associated sequelae, we calculate the ratio of prevented deaths per prevented diagnosis. We consider not only the reduction in diagnoses for the group targeted on the y-axis, but also downstream effects on all other diagnoses, capturing prevented sequelae. We find that there are 0.05 (SD 0.01) prevented deaths per prevented diagnosis due to a reduction of 5\% of new onset malignant neoplasms, suggesting improvements in prevention of malignant neoplasms as the optimal leverage point (Figure \ref{fig:red_deaths} b).    
\subsection{Post-acute Covid-19 sequelae and long-term health impact} 
We next evaluate the impact of the Covid-19 pandemic on diagnosis incidence and all-cause mortality via potential post-acute sequelae.
In the pessimistic scenario (scenario 1), we reduce the HRs by 15\%, and by 40\% in the optimistic scenario (scenario 2). This reflects the potential influence of less severe viral variants and future vaccination campaigns.

\begin{figure*} 
    \centering
    \captionsetup{width=.95\textwidth}
    \includegraphics[width=.8\textwidth]{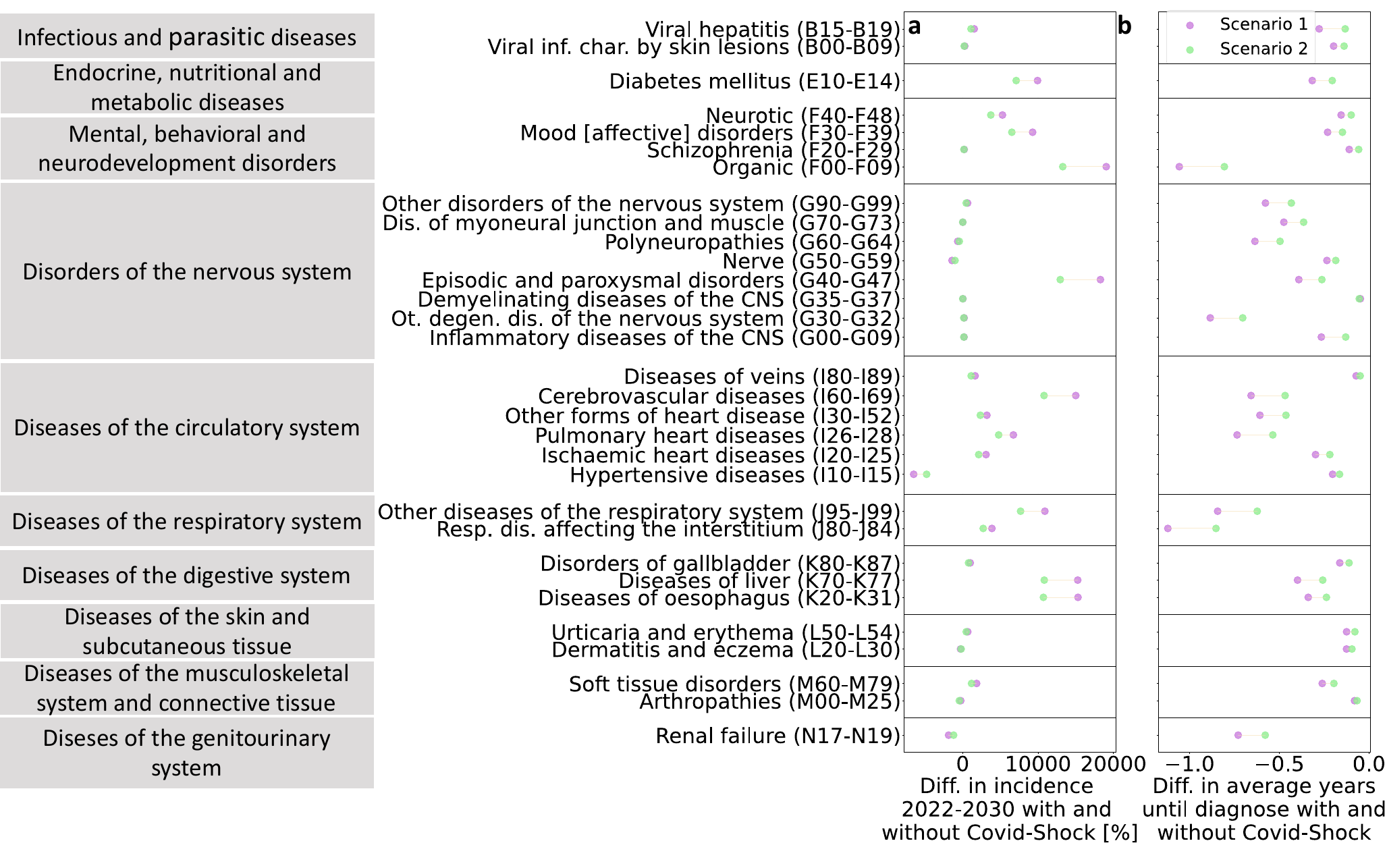}
    \caption{Projected impact of post-COVID-19 sequelae under two scenarios. (a)
    Difference in cumulative diagnoses (2022–2030) for two post-COVID-19 scenarios relative to a no-shock baseline. Scenario 1 assumes a pessimistic trajectory with a 15 $\%$ reduction of the HR of the different diagnosis groups in the y-axis, scenario 2 assumes an optimistic scenario with a 40 $\%$ reduction.Sequelae primarily affect the respiratory and digestive systems. (b) Difference in average time to diagnosis. Negative values indicate earlier onset due to COVID-19 sequelae. Earlier diagnoses are most pronounced for disorders of the respiratory, nervous, and circulatory systems, as well as organic mental disorders (F00–F09), diagnosed up to one year earlier.}
    \label{fig:cs}
\end{figure*}

We analyze the difference in the total number of diagnoses accumulated over nine years (all new onset diagnoses from 2022 to 2030) between the Covid-19 Shock and baseline scenarios (Figure \ref{fig:cs}a). Additionally, we examine the average time until diagnosis between the baseline and the Covid-19 Shock scenarios (Figure \ref{fig:cs}b), serving as an indicator of expedited diagnosis.

The most pronounced differences between the baseline and Covid-19 Shock scenarios are observed for sequelae affecting the respiratory system, digestive system, and mental, behavioral, and neurodevelopment disorders. For other respiratory diseases (J95--J99), scenario 1 predicts 10,905 more diagnoses (equivalent to 0.13 diagnoses per 1,000 persons per year, based on the population in 2022\cite{statistics_austria_projected_2022}), while scenario 2 forecasts 7,684 additional diagnoses (0.09 diagnoses per 1,000 persons per year) compared to the baseline scenario. On average, diagnoses occur 0.84 years earlier in scenario 1 and 0.82 years earlier in scenario 2 than in the baseline scenario.

For respiratory diseases affecting the interstitium (J80--J84), scenario 1 shows 3,900 additional diagnoses (0.05 diagnoses per 1,000 persons per year) and scenario 2 2,700 diagnoses (0.03 diagnoses per 1,000 persons per year) more compared to the baseline scenario, with diagnoses occurring 1.12 and 0.85 years earlier, respectively. We find 15,300 more diagnoses (0.19 diagnoses per 1,000 persons per year) of diseases of the liver (K70--K77) in scenario 1 and 10,800 diagnoses more (0.13 diagnoses per 1,000 persons per year) in scenario 2, with diagnoses occurring 0.4 and 0.26 years earlier, respectively.

Organic, including symptomatic, mental disorders demonstrate 19,000 and 13,300 more diagnoses (0.23 and 0.16 diagnoses per 1,000 persons per year) in scenarios 1 and 2, respectively. Diagnoses occur 1.06 and 0.81 years earlier in scenario 1 and 2, respectively.

We also assess the contrast in the number of fatalities recorded during the period spanning 2022 to 2030 between the baseline and the two Covid-19 Shock scenarios (see Figure \ref{fig:cs_deaths}). Notably, there is a rise in deaths attributed to cerebrovascular diseases (I60--I69: 1147-819 deaths), ischaemic heart diseases (I20--I25: 733-538 deaths), other mental and behavioral disorders (F00--F99: 434-291 deaths) and diabetes mellitus (E10--E14: 324-236 deaths) in response to the Covid-19 Shock. In contrast, fatalities from malignant neoplasms decrease in both Covid-19 Shock scenarios. This may be attributed to individuals succumbing to other causes of death earlier in the simulation.

\subsection{Model performance and calibration accuracy}
Figure \ref{fig:cal} represents the incidences of 131 diagnostic groups resulting from both the simulation and the dataset at the end of the calibration period in 2014. We find good agreement between the annual incidences in the data and the simulation with a Pearson correlation coefficient of $0.99$ across all diagnoses.

Figure \ref{fig:mminc} shows a comparative analysis of the dynamics of multimorbidity incidence between the simulation and the dataset. In addition, we present the projection of multimorbidity incidence for the time span 2028-2030, derived from the simulation results. In Figure \ref{fig:mminc}, the number of individuals receiving $n$ diagnoses is expressed as a percentage within each age group. Notably, marginal fluctuations are observed in the evolution of the multimorbidity incidence. 

The model effectively captures the dynamics of multimorbidity among the younger age cohorts (0-59). For the three oldest age groups ($\geq$ 60), deviations from the empirical data become apparent when comparing the results for 2012-2014 between the data and the model.  In particular, the proportion of highly multimorbid individuals (more than ten diagnoses) is overestimated by a few percentage points in the model. In both data and simulation model, we see that the distribution of the age-specific multimorbidity incidence remains fairly robust over time, even in the projections until 2030. 

We find that some diagnoses show a substantial inflation in the discrepancy between data and simulation over time. There are 36 diagnostic groups with an absolute error in 2014 exceeding 5\% of the 2014 incidence and surpassing 500 cases per annum, as well or those for which error increased by more than 1.3-fold between 2010 and 2014. These diagnoses are excluded from further analysis of changes in incidence.

\section{Discussion} \label{sec:outlook}

We generalize compartmental models traditionally used in infectious disease epidemiology to support long-term simulation of chronic diseases and multimorbidity. Our model, the CDTM, incorporates 132 empirically derived multimorbidity clusters and 131 diagnostic groups (ICD-10: A00–N99), enabling population-scale simulation of disease accumulation. The essence of the model is that it can be fully calibrated from actual nationwide observational data. We present a realization of the model based on an extensive dataset of hospital diagnoses in Austria from 1997 to 2014. We simulate demographic changes (births, migration, deaths) and the acquisition of new diagnoses based on a multimorbdity network. Calibrated from 2003 to 2014 data, the model enables forward simulation of changes in the incidence of diseases and their combinations from 2015 to 2030.

To our knowledge, this is the first national-scale model to forecast multimorbidity trajectories across the entire diagnostic spectrum (A00–N99) while incorporating demographic shifts. This comprehensive microsimulation framework allows for realistic projections of the incidence of individual diseases but also quantitative estimation of health trajectory complexity under different scenarios, including prevention and pandemic-related shocks.  

Diseases affecting the nervous, genitourinary, metabolic, and cardiovascular systems, diagnoses more common among individuals above 65, increase in our baseline scenario where we simulate projected demographic changes (forecasted by the national statistics office). In contrast, the diagnosis groups where we observe a decrease in relative incidence, behavioral disorders with onset in childhood (F90-F98) and diseases of appendix (K35-K38) are more common among individuals under 65. These findings align with expectations, further confirming the validity of our model.
 
Furthermore, the CDTM framework provides concrete quantitative estimates for the future development of such multimorbidity clusters under certain conditions. From 2015 to 2030, a 2.7\% increase in transitions from clusters of three or more diagnoses to four or more is expected, translating to 97,500 additional individuals in 2030. This indicates a growing complexity of health conditions, necessitating comprehensive healthcare strategies. 
Relative occurrence increased from 2015 to 2030 in all 20 of the most common multimorbidity clusters, with the strongest growth in patterns involving hypertensive diseases. 18 out of the 20 most common clusters describe disease patterns including hypertensive diseases (I10-I15), among these are also the most pronounced increases from 2015 to 2030 such as a disease pattern describing hypertensive diseases (I10-I15), other forms of heart disease (I30-I52) and renal failure (N17-N19). This finding is in line with existing literature, which consistently identifies hypertension as one of the most prevalent chronic conditions, either in isolation or in combination with other chronic conditions such as diabetes, hyperlipidemia, ischemic heart disease and chronic kidney disease \cite{kirchberger_patterns_2012, salive_multimorbidity_2013, hawthorne_multimorbidity_2023}. These trends align also with the emerging concept of cardiovascular-kidney-metabolic (CKM) syndrome, which highlights the interconnected progression of hypertension, diabetes, renal dysfunction, and cardiovascular disease. The increasing frequency of such combinations in our simulation reflects shared pathophysiological mechanisms such as inflammation, vascular dysfunction, and metabolic dysregulation, formalized under the CKM framework~\cite{Ndumele2023CKM}.

Estimating the costs of multimorbidity is challenging because co-occurring diseases may have non-additive effects on healthcare spending. Based on a systematic review of direct medical costs and our projected increase in multimorbidity clusters, we estimate the following cost increases between 2015 and 2030: We estimate an additional I\$8.7 million (95\% CI: I\$8.5–8.9 million) for hypertensive diseases co-occurring with dorsopathies and an additional I\$29.8 million (29.1–30.5 million) for hypertensive diseases and arthropathies, assuming annual costs of I\$13,300 per patient. We project that hypertension combined with diabetes mellitus will cost an additional \$13.1 million (12.8–13.5 million) and hypertension combined with other forms of heart disease will cost an additional \$37.2 million (36.4–38.0 million). These estimates are based on per-patient annual costs of \$14,300 and \$17,900, respectively~\cite{tran_costs_2022}.
 

The largest differences in the relative occurrence increase between men and women is observed in disease patterns involving renal failure, with men experiencing a larger increase. This aligns with existing literature indicating that while chronic kidney disease (CKD, N18) is more prevalent among women, the combination of CKD with cardiovascular diagnoses is more commonly observed among men \cite{toth-manikowski_sex_2021, mayne_sex_2023}.

The CDTM supports flexible scenario testing. We showcase the use of the model for assessing changes in disease dynamics for scenarios describing decreasing diagnosis dynamics, as well as increasing ones.

First we develop a framework to study the effectiveness of mitigation efforts (such as vaccinations or other preventive measures) in reducing causes of death. We simulate targeted prevention by reducing the probability of diagnosis acquisition for individual conditions by a 5\% reduction in the transition probability for each individual who would be diagnosed in the following timestep with the respective diagnosis. We also reduce the externally imposed incidence of that diagnosis by 5\%. This allows estimation of downstream effects on multimorbidity and mortality. The largest reductions in deaths following incidence reduction occur in diseases affecting the circulatory system (hypertensive diseases (I10-I15), ischaemic heart diseases (I20-I25), cerebrovascular diseases (I60-I69) and diseases of veins (I80-I89)), malignant neoplasms (C00-C97), endocrine, nutritional diseases (metabolic disorders (E70-E90), diabetes mellitus (E10-E14) and disorders of the thyroid gland (E00-E07)) and diseases of the respiratory system (acute respiratory infections (J00-J06)). These findings indicate that targeting these diagnoses can effectively promote healthy aging.

For a detailed discussion on the medical implications of our findings refer to the supplementary information.

We use CDTM to simulate long-term effects of post-acute SARS-CoV-2 sequelae based on published hazard ratios, modeling both optimistic and pessimistic scenarios. Compared to baseline, we project large increases in incidence from 2022 to 2030: other respiratory diseases (43–56\%, 7,700–10,900 cases), interstitial respiratory diseases (44–56\%, 2,700–3,900), liver diseases (32–42\%, 10,800–15,300), and mental disorders (29–36\%, 13,300–19,000). Most of these increases align with sequelae associated with the highest hazard ratios, except for liver disease. This suggests that the increase in relative incidence of liver diseases is not solely a direct effect of the Covid-19 shock but may also be attributed to it being a sequelae of another condition. A study of the costs of multimorbidity found average annual direct medical costs of I\$ 36,800 (36,400-37,300) for respiratory and mental health conditions and I\$ 35,000 (34,900-35,300) for respiratory and heart/vascular conditions \cite{tran_costs_2022}. 
Additionally, individuals are diagnosed with mental and interstitial respiratory disorders 0.8–1.1 years earlier on average under the shock scenarios. This temporal shift suggests greater multimorbidity compression in later life, with potential implications for care demand and disease burden accumulation during critical years of aging.

The present study is subject to several limitations. First, the hospitalization dataset ends in 2014, and therefore does not reflect more recent changes in medical practice, such as new treatments or diagnostic techniques. We partially mitigate this by ensuring diagnosis incidence error does not inflate over time (variation <1.3x during 2010–2014).

Second, the data does not distinguish between patients who required no further care and those lost to follow-up for non-medical reasons. The model only examines cumulative health states. It is however reasonable to assume that, especially severe, chronic diseases affect the health state of a patient on a long-term basis. 
Finally, the dataset was originally collected for billing purposes, which leads to a potential bias in terms of over- or under-representation of diagnoses. 

Demographic inputs (births, deaths, migration) are provided by official statistics, 'Statistics Austria'. The forecasts on number of births, migration and deaths (2022 onwards) are taken from the main scenario of Statistics Austria. 
While total mortality projections are used through 2030, cause-of-death data is only available through 2019. We therefore assume that the age-, sex-, and cause-specific distribution of deaths remains constant after 2019. 

Finally, our COVID-19 Shock scenarios are based on literature-derived hazard ratios from early-pandemic cohort studies~\cite{al-aly_high-dimensional_2021, daugherty_risk_2021, xu_long-term_2022, xie_risks_2022}. These should be considered when interpreting the results. We simulate optimistic and pessimistic COVID-19 sequelae scenarios using literature-derived hazard ratios, assuming one infection per person by 2022 and attenuated risks from vaccination and milder variants. These simulations illustrate potential long-term effects, not precise forecasts. Future studies with more recent data on SARS-CoV-2 sequelae can refine these estimates.

These findings highlight the potential of scenario-based chronic disease modeling in informing long-term health policy. By identifying the conditions that cause the greatest increases in disease complexity and costs, the CDTM provides a basis for prioritizing prevention in aging populations. Future work should expand this framework to include socioeconomic disparities, climate shocks, and health systems data and examine how it can be applied across different countries.

\subsection*{Author Contributions}
PK conceptualized and supervised the project. PK and KL devised the analytic methods. KL carried out the analysis and produced the plots and graphics. KL and PK wrote the first draft of the manuscript. AK-W and ST made critical comments regarding the manuscript. AK-W contributed medical expertise regarding the medical interpretation of the findings and in developing medical hypotheses. 
AK-W, ST, KL and PK conducted reviewing and editing of the manuscript. All authors read and approved the final manuscript.
\subsection*{Data Availability Statement}
This study was conducted using a pseudonymized research dataset only accessible for selected research partners under strict data protection regulations. Data on disease cluster and sex- and age-stratified transition rates between the different clusters is available from the authors upon request.

\subsection*{Competing Interests Statement}
The authors have declared that no competing interests exist.
\subsection*{Code Availability}
The underlying code for this study [and training/validation datasets] is not publicly available but may be made available to qualified researchers on request. 

\clearpage
\printbibliography



\newpage
\section*{Supplementary Information}
\setcounter{figure}{0} 
\renewcommand{\thefigure}{S\arabic{figure}}
\setcounter{table}{0} 
\renewcommand{\thetable}{S\arabic{table}} 

\subsection*{Construction of Multimorbidity Network}
The multimorbidity network underlying the compartmental disease trajectory model (CDTM) is developed by Haug \textit{et al.} (2020) \cite{haug_high-risk_2020} based on the same dataset of hospital diagnoses from 1997-2014 in Austria.\\ To construct the network, Haug \textit{et al.} (2020) extracted from the dataset all diagnoses with codes in the range A00 to N99 (1074 codes) for a specific subset of individuals. This subset consists of individuals who did not receive a diagnosis between 1997 and 2002, but did receive a diagnosis between 2003 and 2014, to ensure that only patients who can be assumed to be healthy are included. The diagnoses are grouped into 131 diagnostic groups defined by the WHO (see table \ref{tab:diseasegroups}). Each health state per patient and year is coded as a binary vector with 131 entries. An entry with a value of 1 indicates that the individual was diagnosed with the respective diagnosis group, while a value of 0 indicates the absence of such a diagnosis. Note that these diagnoses are cumulative, indicating that an individual remains diagnosed with a disease throughout the observation period, as the dataset lacks information on the resolution or recovery of diagnoses over time. A hierarchical clustering algorithm (DIVCLUS-T \cite{divclus}) identified 132 multimorbidity patterns. Each cluster, multimorbidity pattern, contains a set of diseases with which each patient in that cluster was diagnosed (inclusion criteria) and a set of diseases with which each patient in that cluster was not diagnosed (exclusion criteria). Each cluster is assigned a vector of 131 probabilities, with probability 0 for exclusion criteria and probability 1 for inclusion criteria. Disease groups that are neither inclusion nor exclusion criteria have probabilities between 0 and 1. For each pair of clusters, there exists a disease that is an inclusion in one cluster and an exclusion in the other. Since a patient's health status includes all the diseases the patient has been diagnosed with, cluster transitions are only possible in one direction. The clusters, multimorbidity patterns, serve as nodes in a multiplexed network, with directed edges representing transition probabilities between these patterns. Each layer of the network corresponds to a specific age and sex group. The directed edges between two clusters $k$ and $j$ are weighted based on the age ($a$) and sex ($g$) dependent probabilities ($q_{g,a,k,j}$) of transitioning from one cluster to another within a single time step. The nodes and links in this multilayer network form the basis for the states (compartments) and their transitions in the CDTM.

 
\newpage
\footnotesize
\newpage 
\subsection*{Additional Discussion} 
\textbf{Potential prevention mechanisms and medical implications of our findings.}
Globally, hypertension and metabolic syndrome components have become leading risk factors for disability-adjusted life years (DALYs), especially in women \cite{institute_for_health_metrics_and_evaluation_global_nodate}. In Austria, hypertension, smoking and high body mass index are significant risk factors, with many people over 60 suffering from undiagnosed or undertreated hypertension \cite{institute_for_health_metrics_and_evaluation_global_nodate, mills_global_2016, sillars_sex_2020}. This condition increases the risk of serious diseases, including heart failure and kidney disease, especially when combined with obesity and diabetes. Prevention of cardiometabolic disorders through better screening, early diagnosis and treatment of modifiable risk factors such as blood pressure, cholesterol, glucose and body weight is essential \cite{ginsburg_women_2023,the_global_cardiovascular_risk_consortium_global_2023}. Public health efforts to reduce smoking, alcohol consumption, obesity, and infections can significantly reduce the incidence of vascular disease and cancer \cite{bachner_health_2018,ginsburg_women_2023}. New drugs such as GLP-1 RA and SGLT2 inhibitors show promise in reducing hyperglycemia, obesity, cardiovascular disease, and renal failure \cite{singh_gender_2020,michaeli_established_2023}. These observations are consistent with the results of our model, demonstrating that prevention efforts for these diagnoses are critical to promoting healthy aging and reducing multimorbidity and premature mortality. Better screening for kidney disease and improved management of diabetes and vascular problems will further reduce the risk of serious outcomes, especially in older adults.

An interesting observation is that reduction of thyroid diseases would have great impact on reduction of multimorbidity and death rates. 
These disorders can be easily detected and treated in most cases. 
Associations between thyroid function and arrhythmias are well known and a thyroid-cardiac axis has been proposed, which may explain the clinical observation of an association between thyroid function and mortality \cite{muller_minor_2022}. Hyperthyroidism and thyreotoxicosis can dramatically increase mortality risk \cite{paschou_thyroid_2022}. Both hypothyroidism and hyperthyroidism have been associated with an increased risk of cardiovascular morbidity and mortality in epidemiological studies and meta-analyses. There are several possible pathophysiological mechanisms linking thyroid and cardiovascular disorders such as endothelial dysfunction, blood pressure changes, dyslipidemia, and low-grade systemic inflammation \cite{poredos_endocrine_2023}. \\ Venous disorders (I80-I89), like thyroid disorders, are less common causes of death but are prevalent, especially among women, and represent a significant healthcare burden. Our analysis revealed that a reduction of new onsets of diseases of veins leads to a reduction in potentially deadly sequelae. 
In recent decades, there has been a steady increase in thromboembolic events, including pulmonary embolism, with high rates of recurrence and increased risk of mortality \cite{smith_analysis_2016}. There are multiple risk factors such as increased age, obesity, prolonged immobility, fractures, hospitalisations and conditions associated with impaired haemostasis such as cancer, diabetes, autoimmune diseases and infections including Covid-19 and pregnancy. It is therefore plausible that a greater focus on possible preventive measures and the reduction of venous disease could imply a large reduction in healthcare costs and morbidity. \\
\textbf{Differences to previously published papers.}
At the methodological level, we extend the multimorbidity network from Haug \textit{et al.} (2020) \cite{haug_high-risk_2020} by applying age-dependent scaling factors $\alpha$ to the age- and sex-specific transition probabilities. We calibrate the model from 2003 to 2014 and externally assign discrepancies per diagnosis, age, and sex group to generate rates for forecasting. While unadjusted transition probabilities reproduce diagnosis-specific incidence trends during calibration, they fail to accurately reflect multimorbidity dynamics—specifically, the distribution of individuals with two or more diagnoses. Note, that our population cohort differs from the cohort used to construct the network in \cite{haug_high-risk_2020}, which consisted of individuals with no recorded diagnoses in 1997-2002. The population cohort studied here includes these individuals, hence adjustments in the transition rates are necessary. This highlights a core trade-off between preserving detailed disease-specific trends and accurately modeling population-level multimorbidity trajectories.
\newpage
\begin{longtable}{|c|c|c|}

\caption{131 Disease Groups as defined by the WHO.}\\
ID & ICD-10 & Description \\\hline
0 & A00-A09 & Intestinal infectious diseases \\
1 & A15-A19 & Tuberculosis \\
2 & A20-A28 & Certain zoonotic bacterial diseases \\
3 & A30-A49 & Other bacterial diseases \\
4 & A50-A64 & Infections with a predominantly sexual mode of transmission \\
5 & A65-A69 & Other spirochaetal diseases \\
6 & A70-A74 & Other diseases caused by chlamydiae \\
7 & A75-A79 & Rickettsioses \\
8 & A80-A91 & Viral infections of the central nervous system \\
9 & A92-A99 & Arthropod-borne viral fevers and viral haemorrhagic fevers \\
10 & B00-B09 & Viral infections characterized by skin and mucous membrane lesions \\
11 & B15-B19 & Viral hepatitis \\
12 & B20-B24 & Human immunodeficiency virus [HIV] disease \\
13 & B25-B34 & Other viral diseases \\
14 & B35-B49 & Mycoses \\
15 & B50-B64 & Protozoal diseases \\
16 & B65-B83 & Helminthiases \\
17 & B85-B89 & Pediculosis, acariasis and other infestations \\
18 & B90-B94 & Sequelae of infectious and parasitic diseases \\
19 & B95-B97 & Bacterial, viral and other infectious agents \\
20 & B99-B99 & Other infectious diseases \\
21 & C00-C97 & Malignant neoplasms \\
22 & D00-D09 & In situ neoplasms \\
23 & D10-D36 & Benign neoplasms \\
24 & D37-D48 & Neoplasms of uncertain or unknown behaviour \\
25 & D50-D53 & Nutritional anaemias \\
26 & D55-D59 & Haemolytic anaemias \\
27 & D60-D64 & Aplastic and other anaemias \\
28 & D65-D69 & Coagulation defects, purpura and other haemorrhagic conditions \\
29 & D70-D77 & Other diseases of blood and blood-forming organs \\
30 & D80-D89 & Certain disorders involving the immune mechanism \\
31 & E00-E07 & Disorders of thyroid gland \\
32 & E10-E14 & Diabetes mellitus \\
33 & E15-E16 & Other disorders of glucose regulation and pancreatic internal secretion \\
34 & E20-E35 & Disorders of other endocrine glands \\
35 & E40-E46 & Malnutrition \\
36 & E50-E64 & Other nutritional deficiencies \\
37 & E65-E68 & Obesity and other hyperalimentation \\
38 & E70-E90 & Metabolic disorders \\
39 & F00-F09 & Organic, including symptomatic, mental disorders \\
40 & F10-F19 & Mental and behavioral disorders due to psychoactive substance use \\
41 & F20-F29 & Schizophrenia, schizotypal and delusional disorders \\
42 & F30-F39 & Mood [affective] disorders \\
43 & F40-F48 & Neurotic, stress-related and somatoform disorders \\
44 & F50-F59 & Behavioral syndromes associated with physiological disturbances and physical factors \\
45 & F60-F69 & Disorders of adult personality and behaviour \\
46 & F70-F79 & Mental retardation \\
47 & F80-F89 & Disorders of psychological development \\
48 & F90-F98 & Behavioral and emotional disorders with onset usually occurring in childhood and adolescence \\
49 & F99-F99 & Unspecified mental disorder \\
50 & G00-G09 & Inflammatory diseases of the central nervous system \\
51 & G10-G13 & Systemic atrophies primarily affecting the central nervous system \\
52 & G20-G26 & Extrapyramidal and movement disorders \\
53 & G30-G32 & Other degenerative diseases of the nervous system \\
54 & G35-G37 & Demyelinating diseases of the central nervous system \\
55 & G40-G47 & Episodic and paroxysmal disorders \\
56 & G50-G59 & Nerve, nerve root and plexus disorders \\
57 & G60-G64 & Polyneuropathies and other disorders of the peripheral nervous system \\
58 & G70-G73 & Diseases of myoneural junction and muscle \\
59 & G80-G83 & Cerebral palsy and other paralytic syndromes \\
60 & G90-G99 & Other disorders of the nervous system \\
61 & H00-H06 & Disorders of eyelid, lacrimal system and orbit \\
62 & H10-H13 & Disorders of conjunctiva \\
63 & H15-H22 & Disorders of sclera, cornea, iris and ciliary body \\
64 & H25-H28 & Disorders of lens \\
65 & H30-H36 & Disorders of choroid and retina \\
66 & H40-H42 & Glaucoma \\
67 & H43-H45 & Disorders of vitreous body and globe \\
68 & H46-H48 & Disorders of optic nerve and visual pathways \\
69 & H49-H52 & Disorders of ocular muscles, binocular movement, accommodation and refraction \\
70 & H53-H54 & Visual disturbances and blindness \\
71 & H55-H59 & Other disorders of eye and adnexa \\
72 & H60-H62 & Diseases of external ear \\
73 & H65-H75 & Diseases of middle ear and mastoid \\
74 & H80-H83 & Diseases of inner ear \\
75 & H90-H95 & Other disorders of ear \\
76 & I00-I02 & Acute rheumatic fever \\
77 & I05-I09 & Chronic rheumatic heart diseases \\
78 & I10-I15 & Hypertensive diseases \\
79 & I20-I25 & Ischaemic heart diseases \\
80 & I26-I28 & Pulmonary heart disease and diseases of pulmonary circulation \\
81 & I30-I52 & Other forms of heart disease \\
82 & I60-I69 & Cerebrovascular diseases \\
83 & I70-I79 & Diseases of arteries, arterioles and capillaries \\
84 & I80-I89 & Diseases of veins, lymphatic vessels and lymph nodes, not elsewhere classified \\
85 & I95-I99 & Other and unspecified disorders of the circulatory system \\
86 & J00-J06 & Acute upper respiratory infections \\
87 & J09-J18 & Influenza and pneumonia \\
88 & J20-J22 & Other acute lower respiratory infections \\
89 & J30-J39 & Other diseases of upper respiratory tract \\
90 & J40-J47 & Chronic lower respiratory diseases \\
91 & J60-J70 & Lung diseases due to external agents \\
92 & J80-J84 & Other respiratory diseases principally affecting the interstitium \\
93 & J85-J86 & Suppurative and necrotic conditions of lower respiratory tract \\
94 & J90-J94 & Other diseases of pleura \\
95 & J95-J99 & Other diseases of the respiratory system \\
96 & K00-K14 & Diseases of oral cavity, salivary glands and jaws \\
97 & K20-K31 & Diseases of oesophagus, stomach and duodenum \\
98 & K35-K38 & Diseases of appendix \\
99 & K40-K46 & Hernia \\
100 & K50-K52 & Noninfective enteritis and colitis \\
101 & K55-K63 & Other diseases of intestines \\
102 & K65-K67 & Diseases of peritoneum \\
103 & K70-K77 & Diseases of liver \\
104 & K80-K87 & Disorders of gallbladder, biliary tract and pancreas \\
105 & K90-K93 & Other diseases of the digestive system \\
106 & L00-L08 & Infections of the skin and subcutaneous tissue \\
107 & L10-L14 & Bullous disorders \\
108 & L20-L30 & Dermatitis and eczema \\
109 & L40-L45 & Papulosquamous disorders \\
110 & L50-L54 & Urticaria and erythema \\
111 & L55-L59 & Radiation-related disorders of the skin and subcutaneous tissue \\
112 & L60-L75 & Disorders of skin appendages \\
113 & L80-L99 & Other disorders of the skin and subcutaneous tissue \\
114 & M00-M25 & Arthropathies \\
115 & M30-M36 & Systemic connective tissue disorders \\
116 & M40-M54 & Dorsopathies \\
117 & M60-M79 & Soft tissue disorders \\
118 & M80-M94 & Osteopathies and chondropathies \\
119 & M95-M99 & Other disorders of the musculoskeletal system and connective tissue \\
120 & N00-N08 & Glomerular diseases \\
121 & N10-N16 & Renal tubulo-interstitial diseases \\
122 & N17-N19 & Renal failure \\
123 & N20-N23 & Urolithiasis \\
124 & N25-N29 & Other disorders of kidney and ureter \\
125 & N30-N39 & Other diseases of urinary system \\
126 & N40-N51 & Diseases of male genital organs \\
127 & N60-N64 & Disorders of breast \\
128 & N70-N77 & Inflammatory diseases of female pelvic organs \\
129 & N80-N98 & Noninflammatory disorders of female genital tract \\
130 & N99-N99 & Other disorders of the genitourinary system \\
\label{tab:diseasegroups}
\end{longtable}
 \begin{figure}[h]
     \centering
     \includegraphics[width=0.9\linewidth]{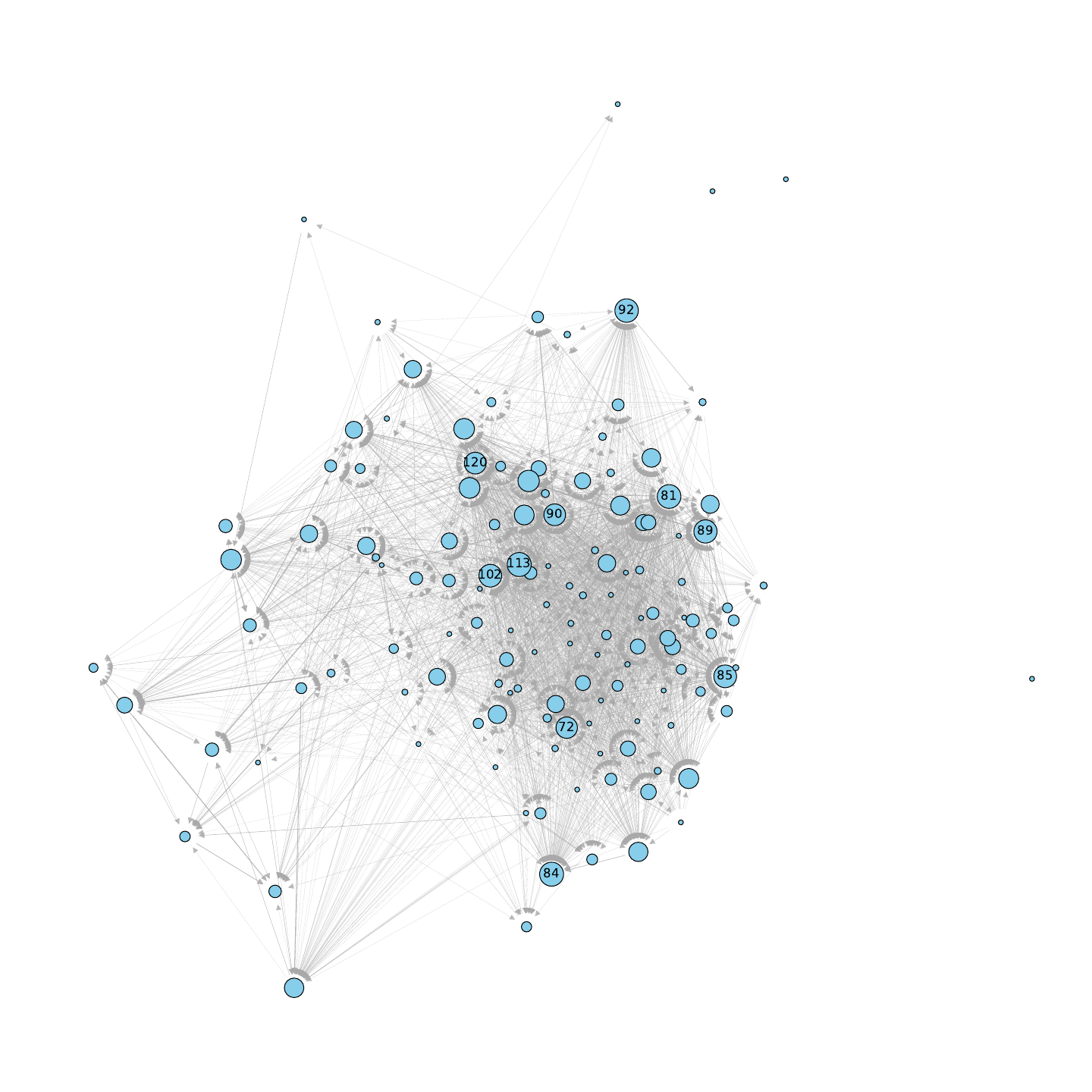}
     \caption{Multimorbidity network for females 50 to 59 years old. Node size indicates in-degree. }
     \label{fig:mmnetwork}
 \end{figure}

 \section{Multimorbidity Cluster}
\label{sec:cluster_criteria}

\renewcommand{\arraystretch}{0.8}
{\tiny
\input{clustertables/cluster_tables/cluster_1}
\FloatBarrier
\input{clustertables/cluster_tables/cluster_2}
\FloatBarrier
\input{clustertables/cluster_tables/cluster_3}
\FloatBarrier
\input{clustertables/cluster_tables/cluster_4}
\FloatBarrier
\input{clustertables/cluster_tables/cluster_5}
\FloatBarrier
\input{clustertables/cluster_tables/cluster_6}
\FloatBarrier
\input{clustertables/cluster_tables/cluster_7}
\FloatBarrier
\input{clustertables/cluster_tables/cluster_8}
\FloatBarrier
\input{clustertables/cluster_tables/cluster_9}
\FloatBarrier
\input{clustertables/cluster_tables/cluster_10}
\FloatBarrier
\input{clustertables/cluster_tables/cluster_11}
\FloatBarrier
\input{clustertables/cluster_tables/cluster_12}
\FloatBarrier
\input{clustertables/cluster_tables/cluster_13}
\FloatBarrier
\input{clustertables/cluster_tables/cluster_14}
\FloatBarrier
\input{clustertables/cluster_tables/cluster_15}
\FloatBarrier
\input{clustertables/cluster_tables/cluster_16}
\FloatBarrier
\input{clustertables/cluster_tables/cluster_17}
\FloatBarrier
\input{clustertables/cluster_tables/cluster_18}
\FloatBarrier
\input{clustertables/cluster_tables/cluster_19}
\FloatBarrier
\input{clustertables/cluster_tables/cluster_20}
\FloatBarrier
\input{clustertables/cluster_tables/cluster_21}
\FloatBarrier
\input{clustertables/cluster_tables/cluster_22}
\FloatBarrier
\input{clustertables/cluster_tables/cluster_23}
\FloatBarrier
\input{clustertables/cluster_tables/cluster_24}
\FloatBarrier
\input{clustertables/cluster_tables/cluster_25}
\FloatBarrier
\input{clustertables/cluster_tables/cluster_26}
\FloatBarrier
\input{clustertables/cluster_tables/cluster_27}
\FloatBarrier
\input{clustertables/cluster_tables/cluster_28}
\FloatBarrier
\input{clustertables/cluster_tables/cluster_29}
\FloatBarrier
\input{clustertables/cluster_tables/cluster_30}
\FloatBarrier
\input{clustertables/cluster_tables/cluster_31}
\FloatBarrier
\input{clustertables/cluster_tables/cluster_32}
\FloatBarrier
\input{clustertables/cluster_tables/cluster_33}
\FloatBarrier
\input{clustertables/cluster_tables/cluster_34}
\FloatBarrier
\input{clustertables/cluster_tables/cluster_35}
\FloatBarrier
\input{clustertables/cluster_tables/cluster_36}
\FloatBarrier
\input{clustertables/cluster_tables/cluster_37}
\FloatBarrier
\input{clustertables/cluster_tables/cluster_38}
\FloatBarrier
\input{clustertables/cluster_tables/cluster_39}
\FloatBarrier
\input{clustertables/cluster_tables/cluster_40}
\FloatBarrier
\input{clustertables/cluster_tables/cluster_41}
\FloatBarrier
\input{clustertables/cluster_tables/cluster_42}
\FloatBarrier
\input{clustertables/cluster_tables/cluster_43}
\FloatBarrier
\input{clustertables/cluster_tables/cluster_44}
\FloatBarrier
\input{clustertables/cluster_tables/cluster_45}
\FloatBarrier
\input{clustertables/cluster_tables/cluster_46}
\FloatBarrier
\input{clustertables/cluster_tables/cluster_47}
\FloatBarrier
\input{clustertables/cluster_tables/cluster_48}
\FloatBarrier
\input{clustertables/cluster_tables/cluster_49}
\FloatBarrier
\input{clustertables/cluster_tables/cluster_50}
\FloatBarrier
\input{clustertables/cluster_tables/cluster_51}
\FloatBarrier
\input{clustertables/cluster_tables/cluster_52}
\FloatBarrier
\input{clustertables/cluster_tables/cluster_53}
\FloatBarrier
\input{clustertables/cluster_tables/cluster_54}
\FloatBarrier
\input{clustertables/cluster_tables/cluster_55}
\FloatBarrier
\input{clustertables/cluster_tables/cluster_56}
\FloatBarrier
\input{clustertables/cluster_tables/cluster_57}
\FloatBarrier
\input{clustertables/cluster_tables/cluster_58}
\FloatBarrier
\input{clustertables/cluster_tables/cluster_59}
\FloatBarrier
\input{clustertables/cluster_tables/cluster_60}
\FloatBarrier
\input{clustertables/cluster_tables/cluster_61}
\FloatBarrier
\input{clustertables/cluster_tables/cluster_62}
\FloatBarrier
\input{clustertables/cluster_tables/cluster_63}
\FloatBarrier
\input{clustertables/cluster_tables/cluster_64}
\FloatBarrier
\input{clustertables/cluster_tables/cluster_65}
\FloatBarrier
\input{clustertables/cluster_tables/cluster_66}
\FloatBarrier
\input{clustertables/cluster_tables/cluster_67}
\FloatBarrier
\input{clustertables/cluster_tables/cluster_68}
\FloatBarrier
\input{clustertables/cluster_tables/cluster_69}
\FloatBarrier
\input{clustertables/cluster_tables/cluster_70}
\FloatBarrier
\input{clustertables/cluster_tables/cluster_71}
\FloatBarrier
\input{clustertables/cluster_tables/cluster_72}
\FloatBarrier
\input{clustertables/cluster_tables/cluster_73}
\FloatBarrier
\input{clustertables/cluster_tables/cluster_74}
\FloatBarrier
\input{clustertables/cluster_tables/cluster_75}
\FloatBarrier
\input{clustertables/cluster_tables/cluster_76}
\FloatBarrier
\input{clustertables/cluster_tables/cluster_77}
\FloatBarrier
\input{clustertables/cluster_tables/cluster_78}
\FloatBarrier
\input{clustertables/cluster_tables/cluster_79}
\FloatBarrier
\input{clustertables/cluster_tables/cluster_80}
\FloatBarrier
\input{clustertables/cluster_tables/cluster_81}
\FloatBarrier
\input{clustertables/cluster_tables/cluster_82}
\FloatBarrier
\input{clustertables/cluster_tables/cluster_83}
\FloatBarrier
\input{clustertables/cluster_tables/cluster_84}
\FloatBarrier
\input{clustertables/cluster_tables/cluster_85}
\FloatBarrier
\input{clustertables/cluster_tables/cluster_86}
\FloatBarrier
\input{clustertables/cluster_tables/cluster_87}
\FloatBarrier
\input{clustertables/cluster_tables/cluster_88}
\FloatBarrier
\input{clustertables/cluster_tables/cluster_89}
\FloatBarrier
\input{clustertables/cluster_tables/cluster_90}
\FloatBarrier
\input{clustertables/cluster_tables/cluster_91}
\FloatBarrier
\input{clustertables/cluster_tables/cluster_92}
\FloatBarrier
\input{clustertables/cluster_tables/cluster_93}
\FloatBarrier
\input{clustertables/cluster_tables/cluster_94}
\FloatBarrier
\input{clustertables/cluster_tables/cluster_95}
\FloatBarrier
\input{clustertables/cluster_tables/cluster_96}
\FloatBarrier
\input{clustertables/cluster_tables/cluster_97}
\FloatBarrier
\input{clustertables/cluster_tables/cluster_98}
\FloatBarrier
\input{clustertables/cluster_tables/cluster_99}
\FloatBarrier
\input{clustertables/cluster_tables/cluster_100}
\FloatBarrier
\input{clustertables/cluster_tables/cluster_101}
\FloatBarrier
\input{clustertables/cluster_tables/cluster_102}
\FloatBarrier
\input{clustertables/cluster_tables/cluster_103}
\FloatBarrier
\input{clustertables/cluster_tables/cluster_104}
\FloatBarrier
\input{clustertables/cluster_tables/cluster_105}
\FloatBarrier
\input{clustertables/cluster_tables/cluster_106}
\FloatBarrier
\input{clustertables/cluster_tables/cluster_107}
\FloatBarrier
\input{clustertables/cluster_tables/cluster_108}
\FloatBarrier
\input{clustertables/cluster_tables/cluster_109}
\FloatBarrier
\input{clustertables/cluster_tables/cluster_110}
\FloatBarrier
\input{clustertables/cluster_tables/cluster_111}
\FloatBarrier
\input{clustertables/cluster_tables/cluster_112}
\FloatBarrier
\input{clustertables/cluster_tables/cluster_113}
\FloatBarrier
\input{clustertables/cluster_tables/cluster_114}
\FloatBarrier
\input{clustertables/cluster_tables/cluster_115}
\FloatBarrier
\input{clustertables/cluster_tables/cluster_116}
\FloatBarrier
\input{clustertables/cluster_tables/cluster_117}
\FloatBarrier
\input{clustertables/cluster_tables/cluster_118}
\FloatBarrier
\input{clustertables/cluster_tables/cluster_119}
\FloatBarrier
\input{clustertables/cluster_tables/cluster_120}
\FloatBarrier
\input{clustertables/cluster_tables/cluster_121}
\FloatBarrier
\input{clustertables/cluster_tables/cluster_122}
\FloatBarrier
\input{clustertables/cluster_tables/cluster_123}
\FloatBarrier
\input{clustertables/cluster_tables/cluster_124}
\FloatBarrier
\input{clustertables/cluster_tables/cluster_125}
\FloatBarrier
\input{clustertables/cluster_tables/cluster_126}
\FloatBarrier
\input{clustertables/cluster_tables/cluster_127}
\FloatBarrier
\input{clustertables/cluster_tables/cluster_128}
\FloatBarrier
\input{clustertables/cluster_tables/cluster_129}
\FloatBarrier
\input{clustertables/cluster_tables/cluster_130}
\FloatBarrier
\input{clustertables/cluster_tables/cluster_131}
\FloatBarrier}
 
 \begin{figure}[h]
    \centering
    \includegraphics[width=.7\textwidth]{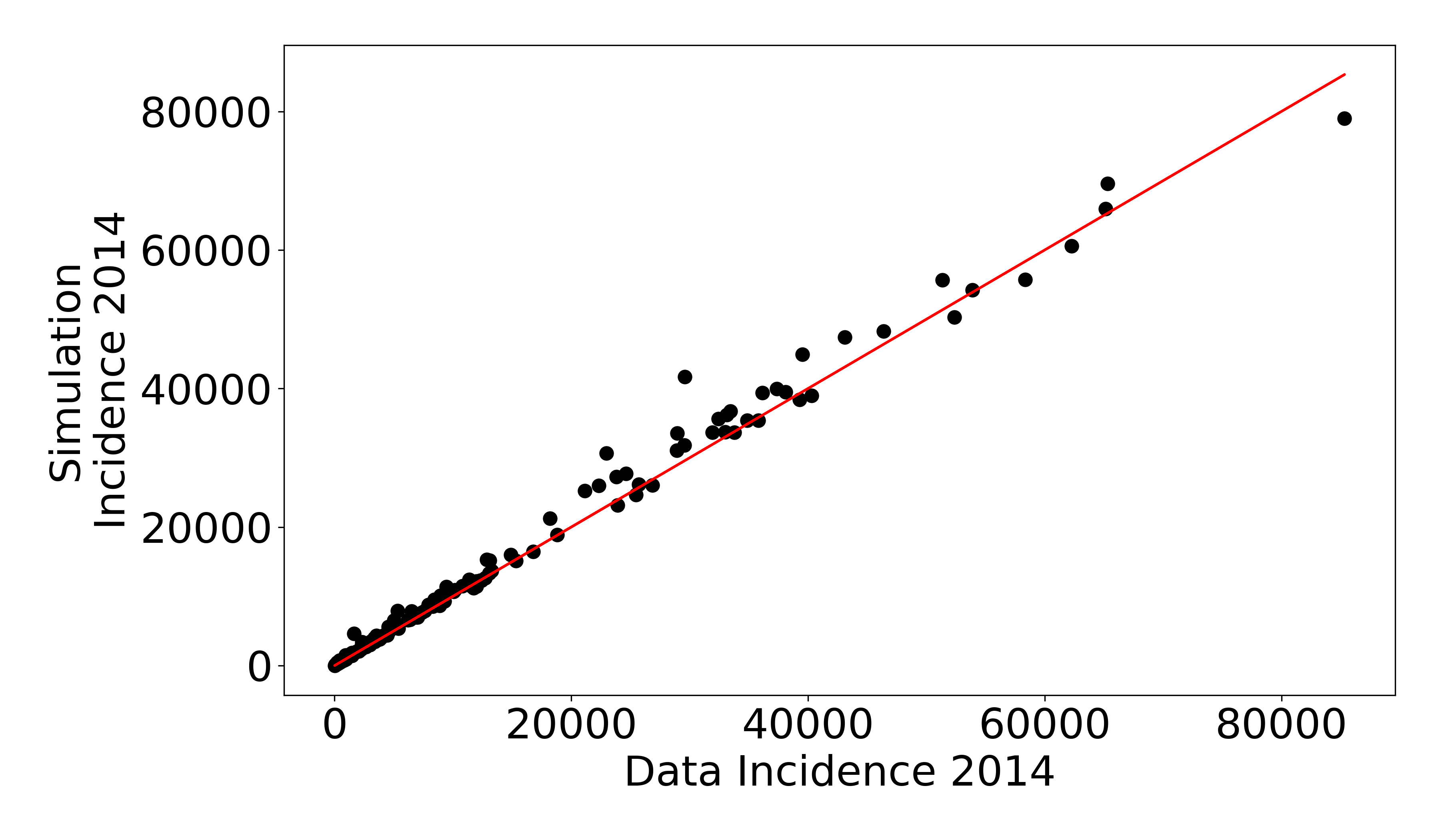}
    \caption{Incidence of the 131 diagnosis groups in simulation year 2014 and incidence of the 131 diagnosis groups in the data, 2014. }
    \label{fig:cal}
\end{figure}
 \begin{figure*} 
    \centering
    \includegraphics[width=.8\textwidth]{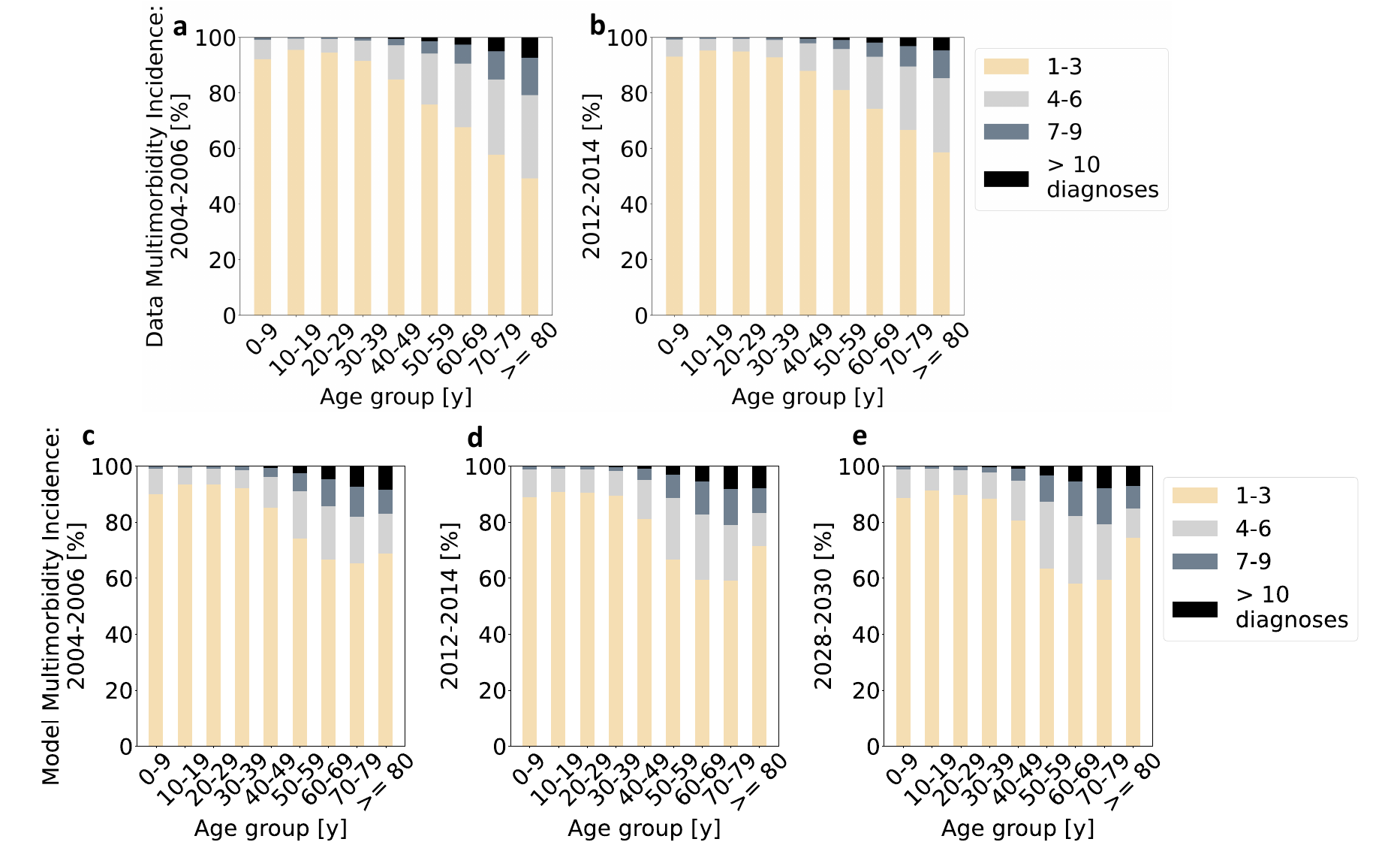}
    \caption{Multimorbidity Incidence observed in the dataset and the simulation results. Each figure shows the proportion of individuals with k diagnoses (ranging from 1-3 to more than 10 diagnoses) among all individuals being diagnosed with at least one diagnosis in the respective age groups. The figures a) and b) in the top panel show the incidence of multimorbdity in the dataset for the periods 2004-2006 and 2012-2014, respectively. In the bottom panel, the figure c), d) and e) show the multimorbidity incidence in the simulation result for the time periods 2004-2006, 2012-2014 and 2028-2030, respectively. Comparing a) and c), and b) and d) shows that the model represents the multimorbdity trends well in the younger age groups (from 0 to 59). The model shows deviations from the resulting incidence in the data in respect to the last three age groups, especially for individuals aged 80 years or older. The proportion of individuals in the simulation result with 1-3 diagnoses in the first six age groups ranges from 93 $\%$ for individuals aged 20-29 to 74 $\%$ for individuals aged 50-59 for the time period 2004-2006. In the dataset, we observe 95 $\%$ for individuals aged 20-29 to 76 $\%$ for individuals aged 50-59. The mean absolute deviation for all k diagnoses in these age groups is 1$\%$. For individuals aged 60 years or older the proportion of individuals with 1-3 diagnoses ranges from 66 $\%$ to 68 $\%$ in the simulation result and 68 $\%$ to 50 $\%$ in the dataset. The mean absolute deviation for all k diagnoses in these age groups is 6 $\%$. The simulation yields among the first six age groups a range of proportion of individuals with 1-3 diagnoses from 91 $\%$ (20-29 years) to 68 $\%$ (50-59 years) in the time period 2014-2016. In the dataset we observe 95 $\%$ (20-29 years) to 81 $\%$ (50-59 years) for the same timeframe. The mean absolute deviation for all k diagnoses in these age groups is 3 $\%$. For the last three age groups the simulation results in a proportion of individuals with 1-3 diagnoses ranging from 61 $\%$ (60-69 years) to 71 $\%$ (>= 80 years) whereas in the data we find a range from 74 $\%$ (60-69 years) to 59 $\%$ (>= 80 years). The mean absolute deviation for all k diagnoses in these age groups is 6 $\%$. }
    \label{fig:mminc}
\end{figure*}
 \begin{figure*} 
    \centering
    \includegraphics[width=.8\textwidth]{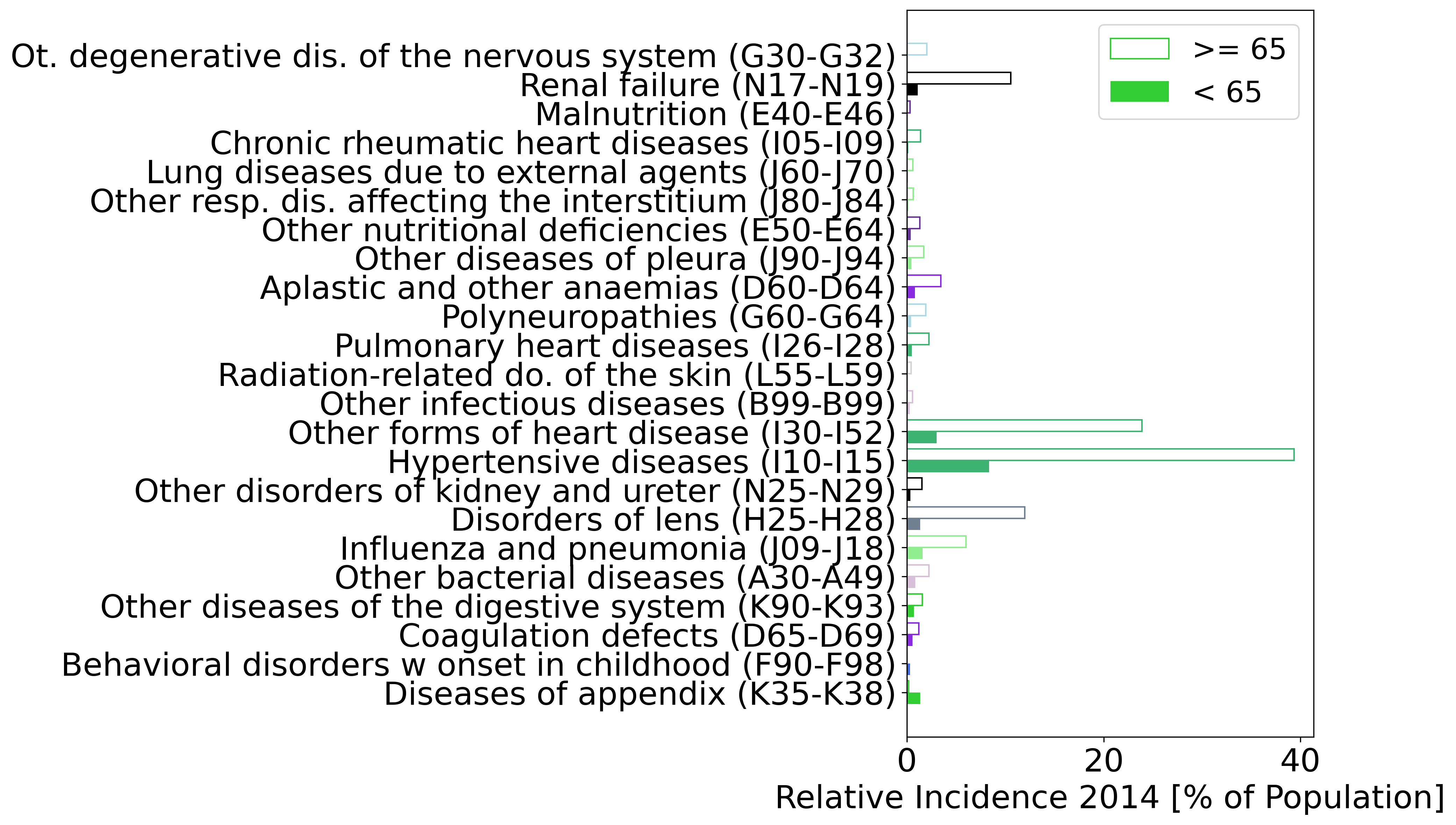}
    \caption{Relative incidence in 2014 (from dataset) of the diagnoses shown in Figure 2. Relative to number of individuals diagnosed in that year (stratified by younger or older than 65). }
    \label{fig:incage}
\end{figure*}
\begin{table*}
    \centering
    \footnotesize
    \begin{tabular}{c|c|c|c}
         Disease & Disease Group & Hazard & Source  \\
         & & Ratio &\\\hline
         Anxiety Disorder & Neurotic, stress-related and somatoform disorders (F40-F48)& 1.38 & \cite{xu_long-term_2022}\\
         Depressive Disorder & Mood [affective] disorders (F30-F39)& 1.44 & \cite{xu_long-term_2022} \\
         Ischemic stroke & Cerebrovascular diseases (I60-I69)& 1.5 &  \cite{xu_long-term_2022}\\
         Myocarditis & Other forms of heart disease (I30-I52)& 5.38 &  \cite{al-aly_high-dimensional_2021}\\
         Myocardial Infarction & Ischaemic heart diseases (I20-I25)& 1.63 &  \cite{al-aly_high-dimensional_2021}\\
         Angina & Ischaemic heart diseases (I20-I25) & 1.52 &  \cite{al-aly_high-dimensional_2021}\\
         Heart Failure & Other forms of heart disease (I30-I52)& 1.62 &  \cite{al-aly_high-dimensional_2021}\\
         Non-ischemic Cardiomyopthy & Other forms of heart disease (I30-I52)& 2.45 &  \cite{al-aly_high-dimensional_2021}\\
         Cardiac Arrest & Other forms of heart disease (I30-I52)& 2.43 &  \cite{al-aly_high-dimensional_2021} \\
         Pulmonary Embolism & Pulmonary heart disease and & 2.93 &  \cite{al-aly_high-dimensional_2021}\\
          &  diseases of pulmonary circulation (I26-I28)& &  \\
         Migraine & Episodic and paroxysmal disorders (G40-G47)& 1.21 &  \cite{xu_long-term_2022}\\
         Peripheral Neuropathy & Polyneuropathies and other disorders & 1.34 &\cite{xu_long-term_2022} \\
         &  of the peripheral nervous system (G60-G64)& &\\
         Encephalopathy & Other disorders of the nervous system (G90-G99)& 1.82 & \cite{daugherty_risk_2021}\\
         Seizure & Episodic and paroxysmal disorders (G40-G47)& 1.8 & \cite{xu_long-term_2022}\\
         Dementia & Other degenerative diseases of the nervous system (G30-G32)& 4.00 & \cite{daugherty_risk_2021}\\
         & Organic, including symptomatic, mental disorders (F00-F09)& & \\
         Type 2 Diabetes & Diabetes mellitus (E10-E14)& 1.4 & \cite{xie_risks_2022} \\
         Liver Test Abnormality & Diseases of liver (K70-K77)\& Diseases of veins, lymphatic vessels and & 1.70 &\cite{daugherty_risk_2021} \\
         & lymph nodes, not elsewhere classified (I80-I89)\& Viral hepatitis (B15-B19)& &\cite{daugherty_risk_2021} \\
         Kidney Injury & Hypertensive diseases (I10-I15)\& Renal failure (N17-N19)& 1.53 &\cite{daugherty_risk_2021} \\
         Respiratory Failure & Other diseases of the respiratory system (J95-J99)& 3.67 &\cite{daugherty_risk_2021} \\
         Interstitial Lung Disease& Other respiratory diseases principally affecting the interstitium (J80-J84)& 7.00 & \cite{daugherty_risk_2021}\\
         Atopic Dermatitis & Dermatitis and eczema (L20-L30)& 1.11 & \cite{daugherty_risk_2021}\\
         Urticaria & Urticaria and erythema (L50-L54)& 1.40 &\cite{daugherty_risk_2021} \\
         Herpesviral vesicular dermatitis  & Viral infections characterized by skin and mucous membrane lesions (B00-B09)& 1.43 & \cite{daugherty_risk_2021}\\
         Sleep Apnea & Episodic and paroxysmal disorders (G40-G47)& 1.75 & \cite{daugherty_risk_2021}\\
         Transient ischemic attacks &Episodic and paroxysmal disorders (G40-G47)&1.62&\cite{xu_long-term_2022}\\
         Hemorrhagic stroke &Cerebrovascular diseases (I60-I69)&2.19&\cite{xu_long-term_2022}\\
         Cerebral venous thrombosis &Cerebrovascular diseases (I60-I69)&2.69&\cite{xu_long-term_2022}\\
         Alzheimer &Other degenerative diseases of the nervous system (G30-G32)&2.03&\cite{xu_long-term_2022}\\
         Dysautonomie&Other disorders of the nervous system (G90-G99)&1.3&\cite{xu_long-term_2022}\\
         Bells palsy&Nerve, nerve root and plexus disorders (G50-G59)&1.48&\cite{xu_long-term_2022}\\
         Stress and adjustment disorder&Neurotic, stress-related and somatoform disorders (F40-F48)&1.39&\cite{xu_long-term_2022}\\
         Psychotic disorder &Schizophrenia, schizotypal and delusional disorders (F20-F29)&1.51&\cite{xu_long-term_2022}\\
         Joint pain&Arthropathies (M00-M25)&1.34&\cite{xu_long-term_2022}\\
         Myalgia&Soft tissue disorders (M60-M79)&1.83&\cite{xu_long-term_2022}\\
         Myopathy&Diseases of myoneural junction and muscle (G70-G73)&2.76&\cite{xu_long-term_2022}\\
         Guillan Barre Syndrome&Polyneuropathies and other disorders&2.16&\cite{xu_long-term_2022}\\
         &of the peripheral nervous system (G60-G64)&&\\
         Transverse Myelitis&Demyelinating diseases of the central nervous system (G35-G37)&1.49&\cite{xu_long-term_2022}\\
         Cholangitis&Disorders of gallbladder, biliary tract and pancreas (K80-K87)&2.02&\cite{xu_long-term_2023}\\
         IBS&Other diseases of intestines (K55-K63)&1.54&\cite{xu_long-term_2023}\\
         Acute Gastritis&Diseases of oesophagus, stomach and duodenum (K20-K31)&1.47&\cite{xu_long-term_2023}\\
         Functional Dyspepsie&Diseases of oesophagus, stomach and duodenum (K20-K31)&1.36&\cite{xu_long-term_2023}\\
         Acute Pancreatitis&Diseases of oesophagus, stomach and duodenum (K20-K31)&1.46&\cite{xu_long-term_2023}\\
         GERD&Diseases of oesophagus, stomach and duodenum (K20-K31)&1.35&\cite{xu_long-term_2023}\\
         PUD&Diseases of oesophagus, stomach and duodenum (K20-K31)&1.62&\cite{xu_long-term_2023}\\
    \end{tabular}
    \caption{Hazard Ratios of diagnoses affected by Covid shock simulation.}
    \label{tab:cshr}
\end{table*}

\begin{figure*} 
    \centering
    \includegraphics[width=1\textwidth]{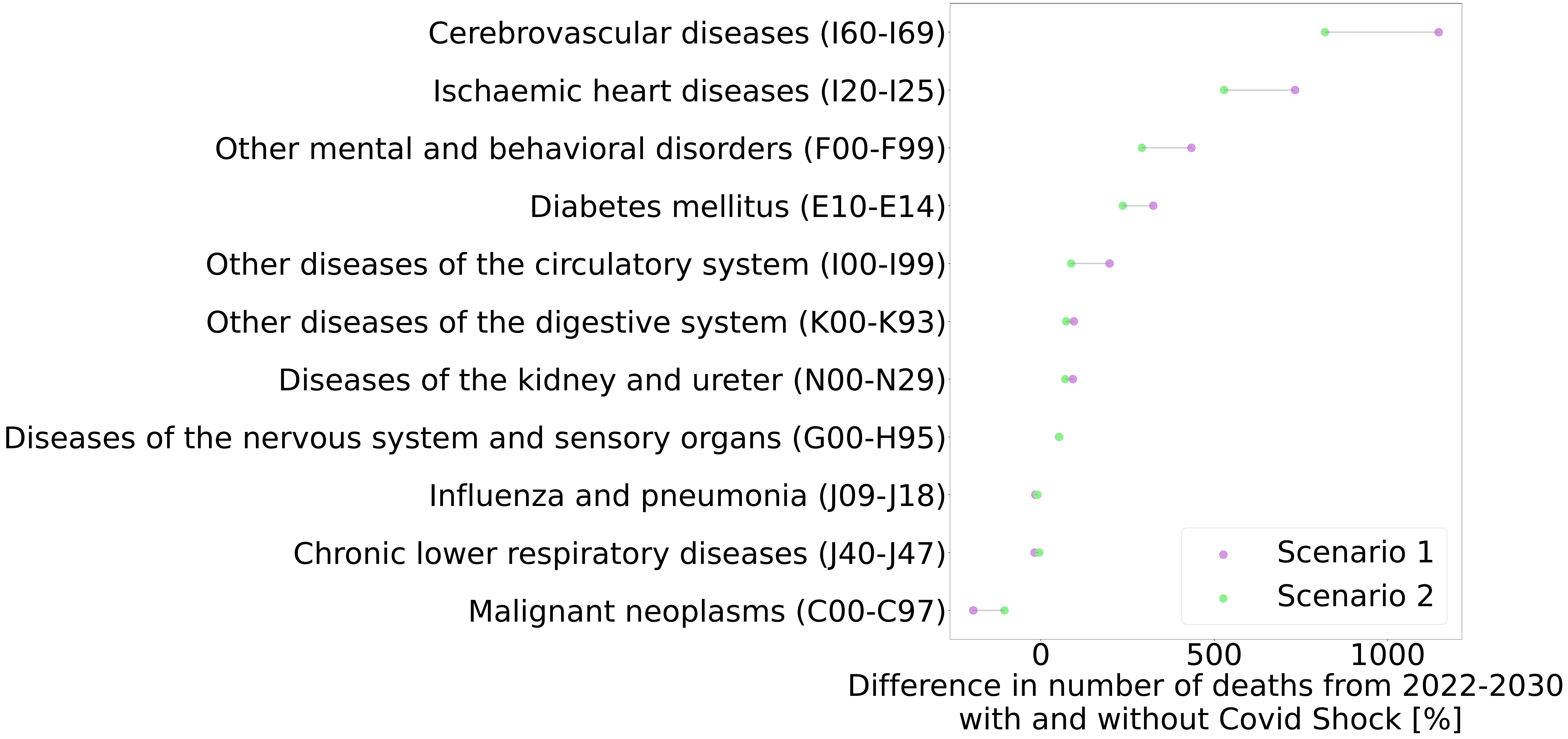}
    \caption{Impact of the Covid Shock Scenarios on death numbers. The figure illustrates the difference in death numbers from 2022 to 2030. There is an increase especially in death causes related to diseases affecting the circulatory system. Deaths due to malignant neoplasms decline across both Covid-19 shock scenarios. This can be attributed to individuals succumbing to other diseases earlier in the simulation. }
    \label{fig:cs_deaths}
\end{figure*}
\begin{figure*} 
    \centering
    \includegraphics[width=1\textwidth]{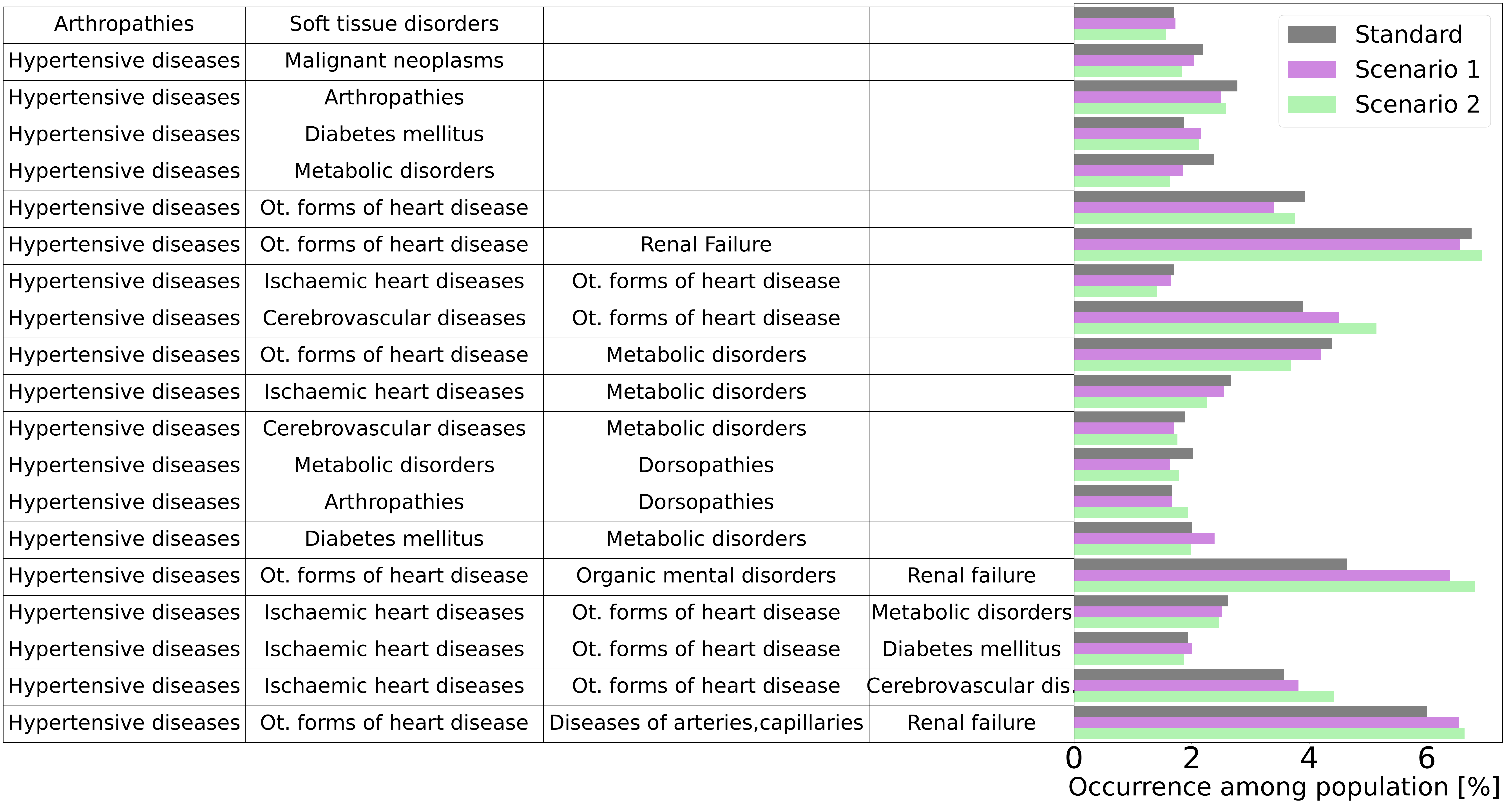}
    \caption{Impact of the Covid Shock Scenarios on multimorbidity pattern occurrence. We show the 20 most common multimorbidity clusters transitioned to in 2030 in the Covid Shock Scenario 1 (pessimistic scenario) and compare the relative occurrence for the standard scenario and both Covid Shock scenarios. }
    \label{fig:cs_cluster}
\end{figure*}
\end{document}

%% file: preamble.tex
\usepackage{amsthm}
\usepackage{mathtools}
\usepackage{physics}
\usepackage{xcolor}
\usepackage{graphicx,caption}
\usepackage[left=23mm,right=13mm,top=35mm,columnsep=15pt]{geometry} 
\usepackage{adjustbox}
\usepackage{placeins}
\usepackage[T1]{fontenc}
\usepackage{lipsum}
\usepackage{csquotes}
\usepackage{hyperref}
\usepackage{url}
\usepackage{longtable}
\usepackage{amssymb}
\usepackage[left]{lineno}